\documentclass[pra,twocolumn]{revtex4}
\usepackage{graphicx}% Include figure files
\usepackage{dcolumn}% Align table columns on decimal point
\usepackage{bm}% bold math
\usepackage{color}
\usepackage{amsmath}
\usepackage{subfig}
\usepackage{hyperref, bbm}
\usepackage[normalem]{ulem}

\newcommand\ot{\otimes}

\newcommand{\<}{\langle}
\renewcommand{\>}{\rangle}

\def\duzomniejsze{<\kern-.7mm<}
\def\duzowieksze{>\kern-.7mm>}
\def\blacksquare{\vrule height 4pt width 3pt depth2pt}

\def\textbf#1{{\bf #1}}
\def\beq{\begin{equation}}
\def\eeq{\end{equation}}
\def\be{\begin{equation}}
\def\ee{\end{equation}}
\def\ben{\begin{eqnarray}}
\def\een{\end{eqnarray}}
\def\beqa{\begin{eqnarray}}
\def\eeqa{\end{eqnarray}}
\def\eea{\end{array}}
\def\bea{\begin{array}}

\usepackage{xspace}
\def\sm{Appendix\xspace}
\def\ma{Main Text\xspace}

\newcommand{\bei}{\begin{itemize}}
\newcommand{\eei}{\end{itemize}}
\newcommand{\bee}{\begin{enumerate}}
\newcommand{\eee}{\end{enumerate}}

\newtheorem{lemma}{Lemma}
\newtheorem{corollary}{Corollary}

\newtheorem{theorem}{Theorem}
\newtheorem{definition}{Definition}
\newtheorem{observation}{Observation}

\def\acal{{\cal A}}

\begin{document}

\title{Quantifying Contextuality}

\author{A. Grudka$^1$, K. Horodecki$^{2,3}$,  M. Horodecki$^{2,4}$, P. Horodecki$^{2,5}$, R. Horodecki$^{2,4}$, P. Joshi$^{4,2}$, W. K\l{}obus$^1$ and A. W\'ojcik$^1$}
\affiliation{$^1$Faculty of Physics, Adam Mickiewicz University, 61-614 Pozna\'n, Poland }
\affiliation{$^2$National Quantum Information Center of Gda\'nsk, 81--824 Sopot, Poland}
\affiliation{$^3$Institute of Informatics, University of Gda\'nsk, 80--952 Gda\'nsk, Poland}
\affiliation{$^4$Institute of Theoretical Physics and Astrophysics, University of Gda\'nsk, 80--952 Gda\'nsk, Poland}
\affiliation{$^5$ Faculty of Applied Physics and Mathematics, Technical University of Gda\'nsk, 80--233 Gda\'nsk, Poland}

\date{\today}% It is always \today, today,

\begin{abstract}
Contextuality is central to both the foundations of quantum theory and to the novel information processing tasks. Although it was recognized before Bell's nonlocality, despite some recent proposals, it still faces a fundamental problem: how to quantify its presence? In this work, we provide a framework for quantifying contextuality. 
We conduct two complementary approaches: (i) bottom-up approach, where we introduce  a communication game, which grasps the phenomenon of contextuality in a quantitative manner;  (ii) top-down approach, where we just postulate two measures - relative entropy of contextuality and contextuality cost, analogous to  existent measures of non-locality (a special case of contextuality). We then match the two approaches, by  showing that the measure emerging from communication scenario turns out to be equal to the relative entropy  of contextuality. We give analytical formulas for the proposed measures for some contextual systems. Furthermore we explore properties of these measures such as monotonicity or additivity.
\end{abstract}

\maketitle

%\begin{multicols}{2}

{\it Introduction:} Non-locality is one of the most interesting manifestations of quantumness of physical systems \cite{Bell}. It exhibits the strength of correlations that comes out of a quantum state when measured independently by distant parties that share it, which is sometimes higher than that coming from classical resources, and can be even higher for super-quantum but non-signaling resources \cite{PR}.
 Nonlocality has been formulated in terms of 'boxes' i.e. families of probability distribution, and has been studied both qualitatively through Bell inequalities as well as quantitatively through measures of non-locality such as cost of non-locality, distillable nonlocality \cite{PR,Brunneretal2011,BrunnerSkrzypczyk,Allcock-wires,Forster} or recently as its (anti)robustness \cite{Joshi-broadcasting}.

There is however another phenomenon known even earlier than Bell's non-locality, called quantum contextuality \cite{Specker-60}. Namely, for certain sets of observables, some of which may be commensurable, their results could not preexist prior to the measurements, or otherwise one would obtain logical contradiction sometimes called as Kochen-Specker paradox \cite{Kochen-Specker}.
In recent years, this phenomenon has been studied in depth. New examples of Kochen-Specker proofs of contextuality has been found \cite{Peres1990-KS,Mermin1990-KS,Mermin-rev} (see also \cite{YuOh,Cabello-13min} and references therein for recent results), and the counterparts of Bell inequalities have been introduced, however in a state independent fashion \cite{Cabello-first-ineq} i.e. that are violated by any quantum state (see also state dependent attempts of \cite{Cabello-ineq-neutrons,Nambu} and \cite{Klyachko,Klyachko-5} for more recent achievements). The fact that quantum theory is contextual has been also treated experimentally \cite{exp-context-photons,exp-context-ions,exp-context-neutrons}, see also \cite{exp-full-context,exp-context-phot-3,exp-context-tests,exp-context-roomt} and references therein for recent results. In fact the phenomenon of non-locality is special case of contextuality: the commensurability relations are provided by the fact that observables are measured on separate systems. Yet it is not vice versa: the phenomenon of contextuality is more basic, as can hold in single partite systems. 

Since the discovery of quantum contextuality there has been a basic problem: {\it How to quantify contextuality?} Only recently there were interesting attempts to quantify contextuality in terms of memory cost \cite{Kleinmann-mem-cost} and the ratio of contextual assignments \cite{Svozil-context-measure}. There were also some measures of non-locality, which is a special case of 
contextuality such as non-locality  cost \cite{PR} and relative entropy of non-locality \cite{vanDam-nl-proofs,Brandao-rel-ent}. In this paper, we propose a program of quantifying contextuality based on two complementary approaches: (i) bottom-up approach, 
where we introduce  a communication game, which grasps the phenomenon of contextuality in a quantitative manner 
(ii) top-down approach, where we just postulate two measures - contextuality cost and relative entropy of contextuality, analogous to  the above mentioned non-locality measures. We then match the two approaches, by 
showing that the measure emerging from communication scenario turns out to be equal to 
the relative entropy  of contextuality.
We further study properties of the measures such as faithfulness, additivity or monotonicity, which are analogous to that of entanglement measures. We also compute it for some systems that possess high symmetries. 

\begin{figure}%[!hbt]
\begin{center}
\includegraphics[width=0.35\textwidth]{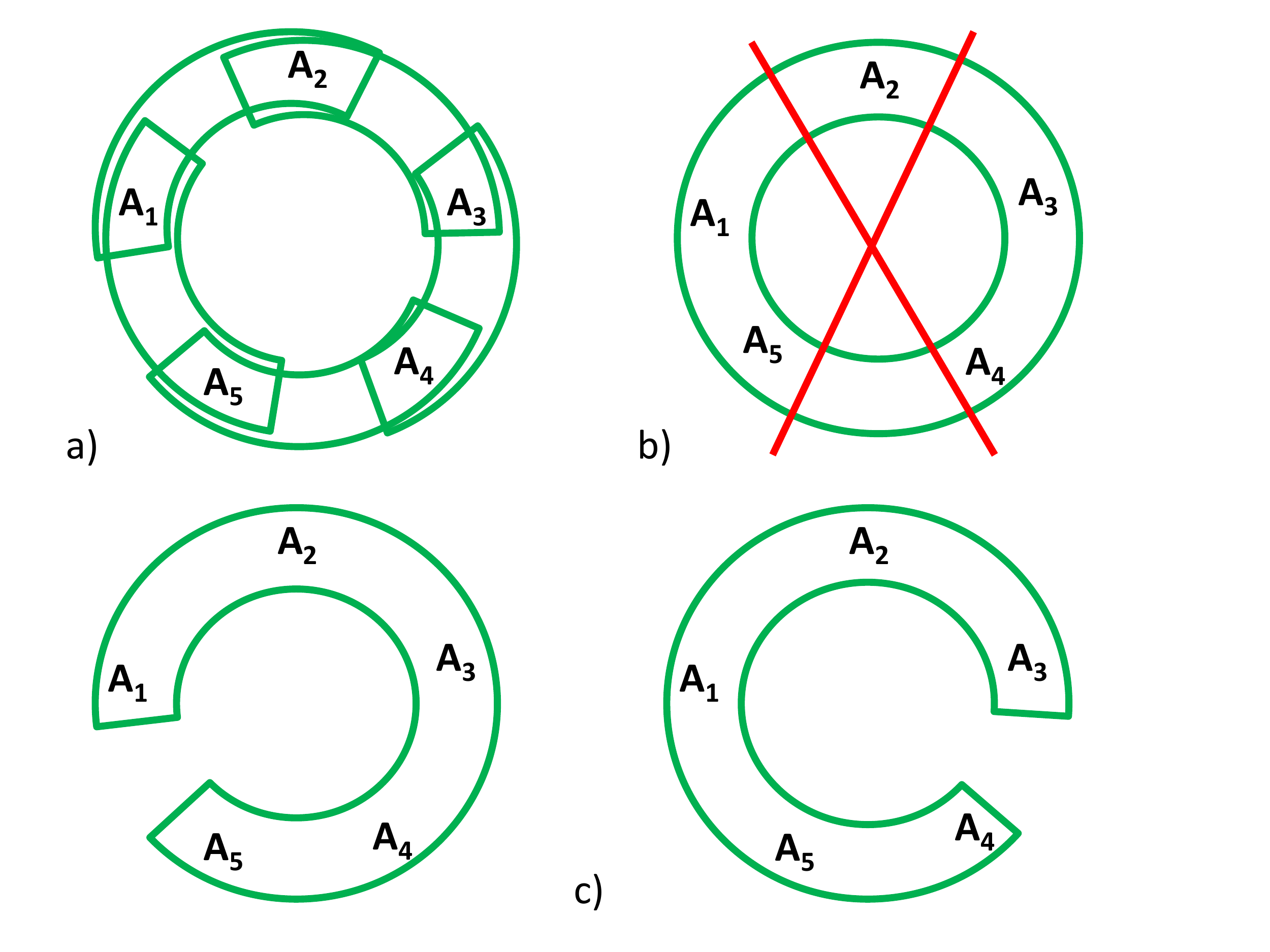}
\end{center}
\caption{Exemplification of contextuality of systems of observables ($A_1,...,A_5$)
a) Contexts (here neighboring $A_i$): observables within each context are jointly measurable, so that we can ascribe joint probability within context.
b) Ascribing single common joint probability distribution which has marginals equal to that ascribed in a) is not possible.
c) Exemplary possible description of the system: by means of two different common joint probability distributions, each of which does not reproduce statistics of some context: the left that of $A_1,A_5$ the right that of $A_3,A_4$.
}
\label{context}
\end{figure}

{\it How to quantify contextuality: }
Quantum contextuality clearly manifests that quantum mechanical world which cannot be described by a joint probability distribution over a single probability space:  there are systems where statistics of observables (some of which are jointly measurable - form a {\it context}), cannot be described by a common joint probability distribution. In other words, joint probability distribution that reproduces statistics of some contexts, see Fig. \ref{context} a), at the same time cannot reproduce statistics of other contexts - see Fig. \ref{context} b). For this reason, if we would like to simulate such a system we need {\it at least two} common joint probability distributions - see Fig. \ref{context} c) where each of them has to fail in reproducing statistics of some context. Thus, for a {\it contextual} systems there are inevitable correlations between the contexts and the common joint probability distributions, while for {\it non-contextual} the "which context information" is inaccessible via the joint probability distribution. We will quantify these correlations by means of mutual information since they vanish iff the system is non-contextual. This quantity will be called the {\it mutual information of contextuality} (MIC). We further show, that it equals another quantity, that can be viewed as an analogue of relative entropy of entanglement, that we call {\it relative entropy of contextuality}. We study properties of this measure, showing it's {\it additivity} for some systems, as well as {\it monotonicity} under some set of operations.
We then compute it for some known systems, developing technique of symmetrization. Finally, we introduce the measure called {\it cost of contextuality} and compute it for some systems.

To formalize the above ideas, we consider a set of observables $V$ some of which are commensurable.
Each set of mutually commensurable observables we call a {\it context}, and assign to it a number $c$. With each context its joint probability distribution over observables that form it, denoted as $g(\lambda_c)$. The set of such contexts $\{g(\lambda_c)\}$ we call a {\it box}. The box is {\it non-contextual} if there exists a joint probability distribution $p(\lambda)$ of all observables in $V$, such that it has marginal distributions on each context $c$ that are equal to $g(\lambda_c)$. Otherwise we call it {\it contextual}. 

\begin{figure}%[!hbt]
\begin{center}
\subfloat[PR box]{\label{figurynka:pr}\includegraphics[width=0.15\textwidth]{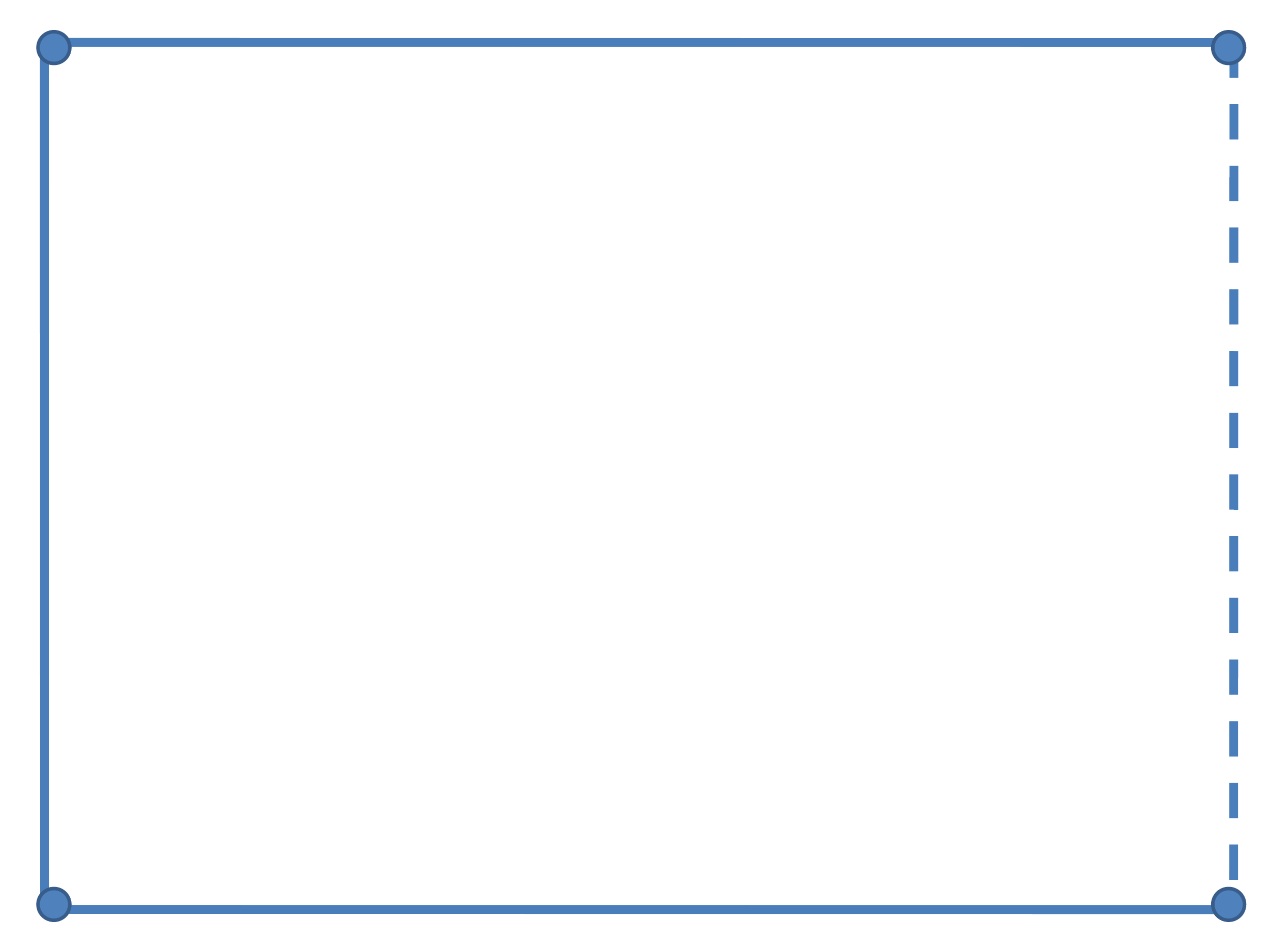}}
\qquad
\subfloat[$CH_{(5)}$ box]
{\label{figurynka:ch}\includegraphics[width=0.18\textwidth]{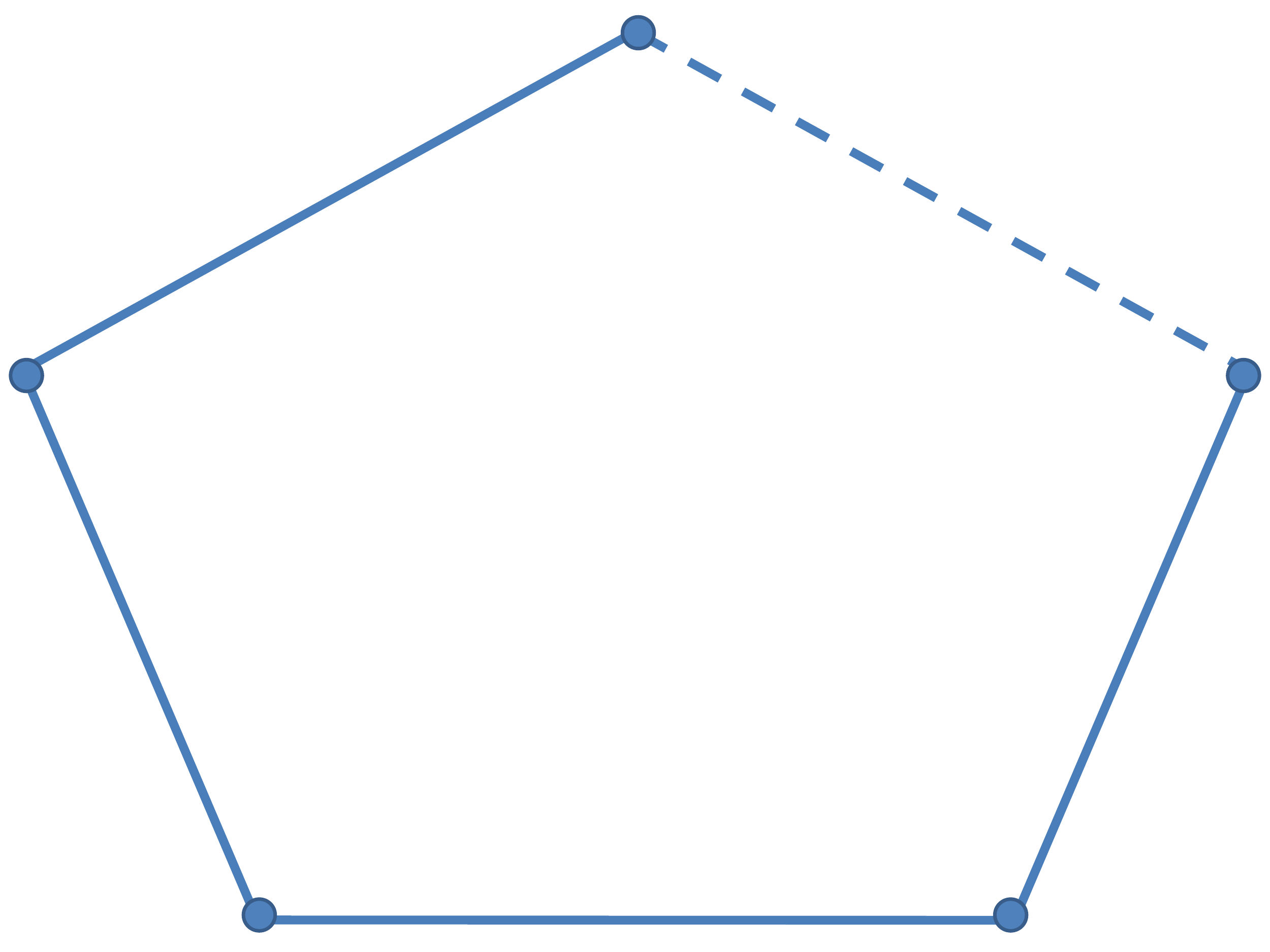}}
\end{center}
\caption{Depiction of the hypergraphs of the Popescu-Rochrlich box (a) and $CH_{(5)}$ box (b). Vertices denotes observables. Each solid line corresponds to a context with fully correlated distribution, dashed one with fully anti-correlated distribution.
}
\label{grafy}
\end{figure}

For illustration, the family of contextual boxes we describe here the so called {\it chain boxes}. The $n$-th chain box, denoted as $CH_{(n)}$ is based on $n$ dichotomic observables $A_1,A_2,...,A_n$, with the $n$ contexts defined as neighboring pairs of observables $A_i,A_{i+1 mod \,n}$. The distributions of these contexts are fully correlated 
%(${1\over 2}$ $00$, ${1\over 2}$ $11$) 
for all but last context and fully anti-correlated 
%(${1\over 2}$ $01$, ${1\over 2}$ $10$) 
for the last one i.e. $A_n,A_1$ \cite{Aruajo-chain}. Note that
$CH_{(4)}$ is the well known Popescu-Rohrlich (PR) box. The boxes which have only two types of distributions of contexts: equally weighted strings with parity 0 and equally weighted bit-strings of parity 1 we call {\it xor-boxes}. The pair: set of observables and set of contexts, form a {\it hypergraph}. The hypergraphs of exemplary xor-boxes \footnote{Note that xor-boxes are those which wins maximally the xor-games i.e. such games for which payoff are only the functions of xor of the output \cite{Oppenheim-Wehner}.} that we consider in the paper are depicted on Fig. \ref{grafy}.

{\it The "which context" game.}-
To formalize introduction of the MIC measure, we consider the following game with three persons: Alice and Bob (the sender and receiver) and Charlie (adversary). Let the parties preagree on some a priori fixed box $B=\{g(\lambda_c)\}$ in hands of Alice. The goal of Alice is to communicate a number of a context $c$ to Bob, through hands of Charlie. To this end she chooses the best probability distribution $\{p(c)\}$, and sends $c$ drawn according to it as a challenge to Charlie. Charlie is bounded to do the following: create a distribution $\acal_c$ over all variables in $V_G$, such that it is compatible with $g(\lambda_c)$ on observables that form context $c$, and send it to Bob. The goal of Charlie is opposite: to disallow communication of $c$ in this way. Bob distinguishes between $\acal_c$'s the best he can. The amount of correlations between Alice and Bob, given Alice's choice of distribution $\{p(c)\}$ achievable in this game is 
\be
I_{\{p(c)\}}(B):=\min_{{{\acal}_c}} I(\sum_{c} p(c)|c\>\<c|\ot {{\acal}_c}),
\label{eq:mutual-form}
\ee
which is the {\it mutual information of contextuality given a priori statistics $\{p(c)\}$ } of a box B.
We use here Dirac notation only for convenience, meaning a classically correlated system of variables ${{\acal}_c}$ correlated with register holding value $c$.
Optimizing over strategies of Alice, we obtain the {\it mutual information of contextuality for a box B } (MIC) i.e. the following quantity:
\be
I_{max}(B) = \sup_{\{p(c)\}} I_{\{p(c)\}}(B).
\label{eq:Imax}
\ee
which reports how much correlations Alice and Bob can obtain in this game.

\begin{figure}%[!hbt]
\begin{center}
\includegraphics[width=0.35\textwidth]{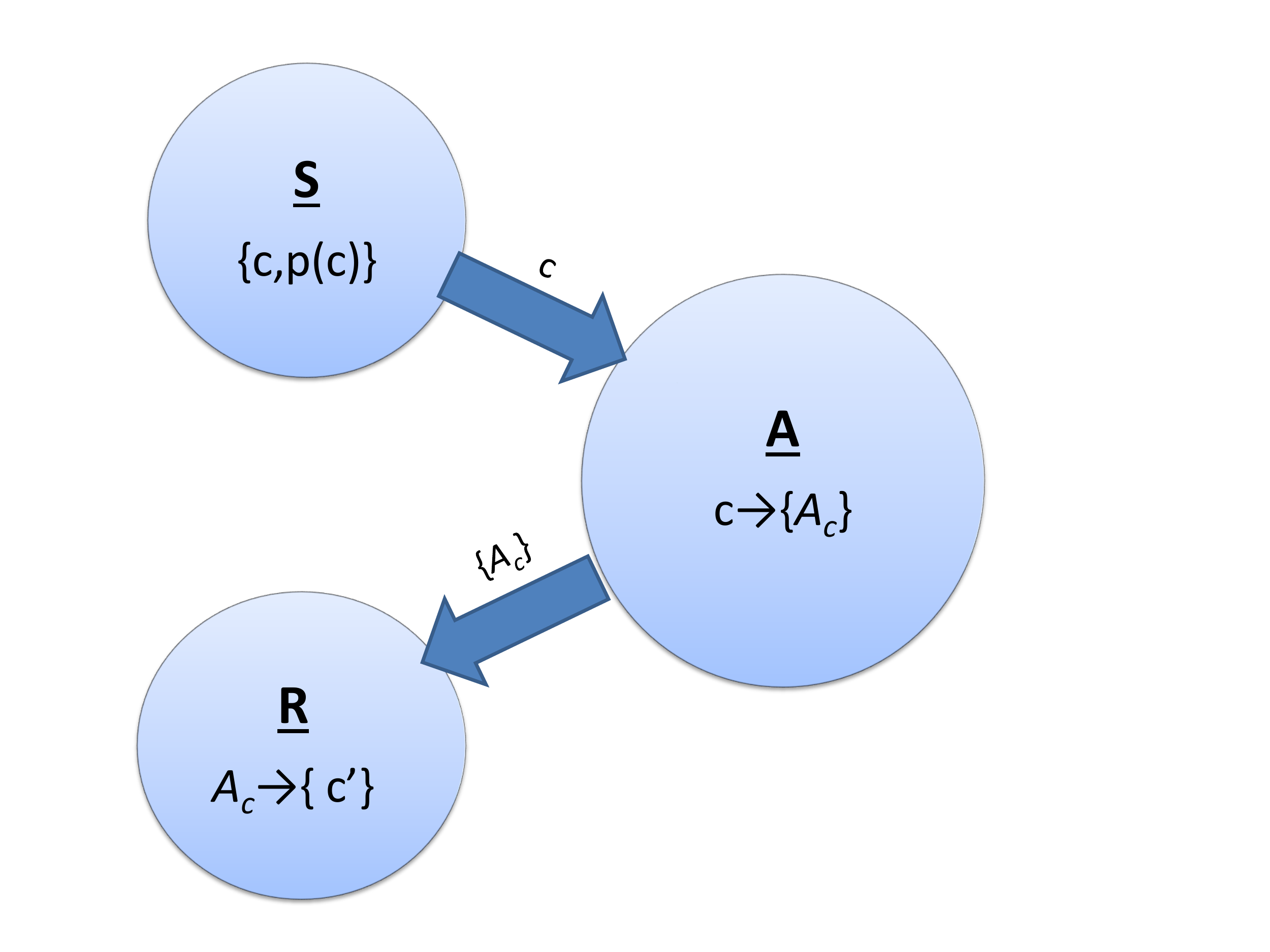}
\end{center}
\caption{ The "which context" game.
\newline The Adversary (A) creates $A_c$ which have context {\it c} as that of a chosen box B such that he minimizes communication from Sender (S) to Receiver (R)}
%\label{grafy}
\end{figure}

We will argue now, that this quantity reports how much contextual is box $B$. Suppose first that $B$ is {\it non-contextual}. Then by definition {\it there exists a single joint probability distribution} $\acal$ over all observables in $V_G$ with marginals $g(\lambda_c)$ on contexts $c$, hence $I_{max}(B)=0$. However in case of {\it contextual} box $B$, by definition Charlie has to use {\it at least two} joint probability distributions of all observables in $V_G$, so that on observables of context $c$, the distribution is $g(\lambda_c)$. Thus, by compactness argument, the value $I_{max}(B)$ is strictly positive.

{\it (Uniform) Relative entropy of contextuality.}-
We introduce now another measure based directly on the notion of relative entropy distance, in analogy to measure of non-locality introduced in \cite{vanDam-nl-proofs}. The first variant, called {\it relative entropy of contextuality} is defined on any box $B = \{g(\lambda_c)\} \in C^{(n)}_G$ as follows:

\be
X_{max}(B):= \sup_{p(c)} \min_{\{p(\lambda)\}} \sum_{c \in E_G} p(c) D(g(\lambda_c) || p(\lambda_c))
\label{eq:xmax}
\ee
where $D(g(\lambda_c)||p(\lambda_c)) = \sum_i g(\lambda_c)_i \log {g(\lambda_c)_i \over p(\lambda_c)_i}$ is the relative entropy distance between distributions $g(\lambda_c)$ and $p(\lambda_c)$ \cite{CoverThomas},\footnote{All logarithms in this paper are binary.}.
The minimization is taken over all distributions $p(\lambda)$ over $\Omega(A_1)\times...\times \Omega(A_k)$ with marginal distribution on context $c$ equal to $p(\lambda_c)$, and supremum is taken over probability distributions $p(c)$ on the set of numbers of contexts $\{1,...,n\}$.

A natural quantity is also the one which does not distinguish the contexts, i.e. instead of maximization we set $p(c) = {1\over n}$ for all $c$:
\be
X_{u}(B):= \min_{p(\lambda)} \sum_{c \in E_G} {1\over n} D(g(\lambda_c) || p(\lambda_c))
\label{eq:xu}
\ee 
where $n$ is number of contexts. We call it the {\it uniform relative entropy of contextuality}. By definition we have $X_{max} \geq X_u$ but in general these measures are not equal since they differ on {\it direct sum} of a contextual and non-contextual boxes (see \sm section \ref{app:not-equal}).

At first it seems that mutual information of contextuality and relative entropy of contextuality are different, and it is not clear how they are related. Interestingly, one can show that they are {\it equal} to each other (see \sm Theorem \ref{thm:equivalence}),
that is:
\be
X_{max} = I_{max}.
\ee

We note here, that $X_u$ and $X_{max}$ (and hence $I_{max}$ according to the above result) are faithful.

{\it Analytical formulas.}- We calculate now the value of $X_u$ and $X_{max}$ for the boxes called {\it isotropic xor-boxes}. To give example of isotropic xor boxes 
we consider here the isotropic chain boxes:
\be
CH^{(n)}_{\alpha} = \alpha CH_{(n)} + (1-\alpha)CH_{(n)}'
\label{eq:fam}
\ee
where $CH_{(n)}'$ is the $CH_{(n)}$ box with correlations and anti-correlations replaced with each other.
We just give idea of how to calculate the (uniform) relative entropy of contextuality for $CH_{(4)}^{\alpha}$ which is isotropic Popescu-Rohrlich box denoted as $PR_{\alpha}$, the detailed proof for other xor-boxes is shown in \sm section \ref{app:computing} and \ref{app:beta}.
The techniques employed are analogous to those used in entanglement theory, including twirling  \cite{Werner1989} as well as using symmetries to compute measures based on distance from the set of separable states \cite{PlenioVedral1998,Rains1999}, and they were applied in the case of  nonlocality e.g. in \cite{Nonsig-theories,Short}. We first compute the value of $X_u$ and then argue, that it equals $X_{max}$ for the isotropic boxes.
The first step is to observe, that for isotropic boxes, in definition of $X_u$
%eq:iso-thm
the minimum can be taken only over those probability distributions $p(\lambda)$ which give rise to an isotropic box, and $p(\lambda_c)$ is marginal of $p(\lambda)$. To show this, we consider $G$ such that $B \in C^{(n)}_G$, and a group of automorphisms of $B$ which can be achieved by operations that transforms $NC_G$ into $NC_G$ i.e. preserve non-contextuality, call it $G_{\mathcal L}$. The idea is to apply to a box $B$ a twirling operation: $B\mapsto \sum_{f\in G_{\mathcal L}} |G_{\mathcal L}|^{-1} f(B)$
where $|G_{\mathcal L}|$ is number of different automorphisms $b_i\circ \pi_i$ which in our case are permutations of contexts $\pi_i$, composed with appropriate negations of outputs of observables $b_i$ (see \sm Theorem \ref{thm:iso-optimal}).

Let us consider an example of $PR_{\alpha}$ box (the other examples of isotropic xor-boxes, follow similar lines, see \sm setion \ref{app:beta}), 
for which
\be
X_{u}(PR_{\alpha}) = \min_{p(\lambda)= PR_{\alpha'}} {1\over 4} \sum_c D(g(\lambda_c)||p(\lambda_c)),
\ee
where $p(\lambda)$ runs over distributions which are from the family of isotropic boxes \cite{Nonsig-theories,Short} that are non-contextual.

\begin{figure}

  \includegraphics[width=0.4\textwidth]{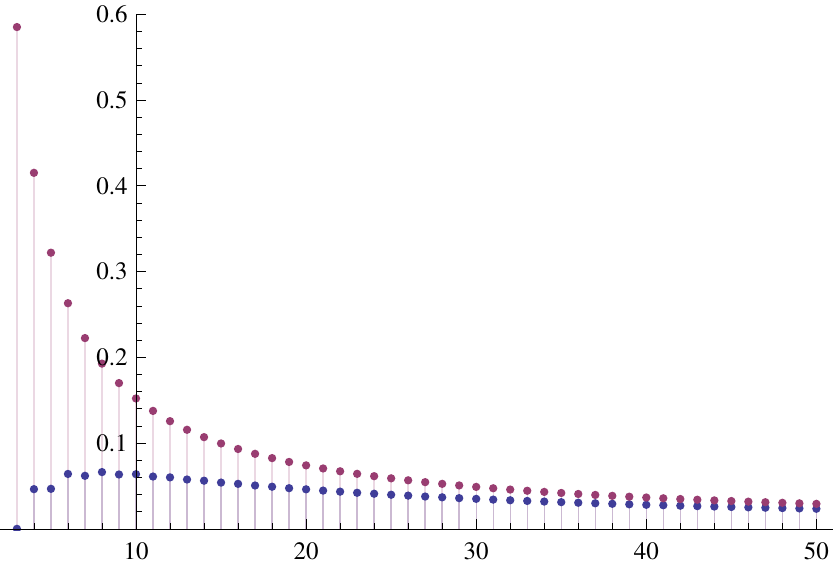}
  \caption{Values of measure $X_u$ for $CH^{(n)}_{\alpha}$ boxes for $3\leq n \leq 50$: maximally contextual boxes (upper points, $\alpha=1$);
  maximally contextual quantum boxes (lower points) with (i) odd $n$  $\alpha=\frac{2\cos(\pi/n)}{1+\cos(\pi/n)}$, (ii) even $n$, $\alpha=(1+\cos(\pi/n))/2$.}   \label{fig:miary_CHn}
\end{figure}

Since any non-contextual box compatible with $G^{(4)}_{CH}$ has to satisfy the inequality which is equivalent to CHSH inequality
${1\over 4} \leq \alpha' \leq {3\over 4}$ (see \sm section \ref{app:beta})
Next step is to observe, that relative entropy does not change under reversible operations such as bit-flip of an output of an observable, (see \sm lemma \ref{lem:one-term}), which gives:
\begin{multline}
X_{u}(PR_{\alpha}) = \min_{{1\over 4}\leq \alpha'\leq {3\over 4}} \nonumber \\
 D(\alpha P^{(2)}_{even} + (1-\alpha)P^{(2)}_{odd}|| \alpha'P^{(2)}_{even} + (1-\alpha')P^{(2)}_{odd}).
\end{multline}
Because all isotropic xor-boxes has the above property, that $X_u(B_{\alpha})$ equals a {\it single} term of relative entropy no matter how many contexts the box $B$ has, we have that for these boxes $X_{max} = X_u$ (see \sm Theorem \ref{thm:xmax-xu}).
It is then easy to show, that for $\alpha \geq {3\over 4}$ there holds
\ben
X_{max}(PR_{\alpha})=X_{u}(PR_{\alpha}) = \log({4\over 3^{\alpha}})  - h(\alpha),
\een
where $h(\alpha) = -\alpha \log\alpha - (1-\alpha)\log(1-\alpha)$ is the binary Shannon entropy. For $\alpha \leq {1\over 4}$, $X_{u}(PR_{\alpha})$
equals the value of $X_u(PR_{(1-\alpha)})$ according to the above equation.
On Fig. \ref{fig:miary_CHn} we present values of measure $X_u$ for chosen chain boxes $CH^{(n)}_{\alpha}$
(quantum ones provided in \cite{Aruajo-chain} and maximally contextual ones).

Analogous considerations gives $X_{max}=X_{u}\approx 0.0467$ for the Klyachko et al. \cite{Klyachko-5} (KCBS) box see \sm subsection \ref{app:computing:K}.

One of the most welcome properties of the measure would be its {\it additivity}.
In \sm Theorem \ref{thm:2add}, we show that for families of isotropic xor-boxes $X_u$ and $X_{max}$ are $2$-copy additive i.e. $X_u(B^{\otimes k})=X_{max}(B^{\otimes k}) = k X(B)$ for $k=2$. For boxes which are extremal within the family of isotropic xor-boxes (such as $CH_{(n)}$, $PM$, $M$) $X_u$ and $X_{max}$ are additive i.e. that the latter statement is true for any natural $k\geq 1$. We conjecture however, that proposed measures are additive for all isotropic xor-boxes.

Another welcome property would be monotonicity of $X_u$ and $X_{max}$ under operations which preserve contextuality. We answer partially this question showing
in \sm subsection \ref{app:props-of-measures} that they are non-increasing under a natural subclass of contextuality preserving operations.

{\it The contextuality cost.}-We would like to note, that there is an obvious way to quantify contextuality using strength of violation of some Kochen-Specker (KS) inequality. This approach however is not universal, since there are boxes that are contextual but do not violate this specific KS-inequality \footnote{ E.g. $PM'$ does not violate inequality (\ref{eq:beta-n-1}) in \sm with $n=6$.}. Thus we demand that our measure of contextuality $X$ should be {\it faithful} i.e. nonzero iff the box is contextual.

Another approach is to base on some known measures of non-locality and define it properly for all (also one-partite) boxes.
This leads us to the {\it contextuality cost}, which we define as follows:
\be
C(B) := \inf \{p\in [0,1]| B = p B_C + (1-p)B_{NC} \}
\ee
where infimum is taken over all decompositions of box $B$ into mixture of some non-contextual box $B_{NC}$ and some contextual box $B_C$.
This measure inherits after nonlocality cost the property that it is  {\it not increasing} under operations that {\it preserve non-contextuality}, which are the operations satisfying the following axioms: (i) transform boxes into boxes
(ii) are linear
(iii) preserve consistency
(iv) transform non-contextual boxes into non-contextual ones. This holds for the same reason for which the anti-robustness of nonlocality is non-increasing under class of {\it locality preserving operations} as it is shown in \cite{Joshi-broadcasting}. We note also that this measure is by definition faithful, and one can easily compute it using linear programming \cite{Brunneretal2011}, it is however not {\it extensive} i.e. is not proportional to dimension of the system.
For the families of isotropic boxes, it can be found analytically namely that $C(PM_{\alpha}) = 6\alpha -5$, $C(M_{\alpha}) = 5\alpha - 4$ and $C(CH^{\alpha}_{(n)})=n\alpha -(n-1)$ (in the same way
as it is shown in \cite{KH-discrim} that $C(PR_{\alpha}) = 4\alpha -3$).

{\it Conclusions.}-We have proposed a framework to quantify contextuality. In particular we have  introduced measures of state dependent/independent contextuality which are valid for both the single and many party scenarios.  Our approach can be developed in different ways. First, one can define analogous measures to $X_u$ and $X_{max}$ setting variational distance in place of relative entropy. 

One can also consider a measure defined as $\min_{{{\acal}_c}} \sup_{p(c)} I(\sum_c p(c)|c\>\<c|\ot {{\acal}_c})$, i.e. with changed order of min and sup in (\ref{eq:Imax}) which for non-local boxes has been studied in \cite{vanDam-nl-proofs}. This measure  have  more communicational meaning than $X_{max}$, it is minimal capacity of the channel from Sender to Receiver under Adversary's attack.
Note, that another way of defining relative entropy of contextuality, would be to consider a quantity defined on a box $B$ compatible with graph $G$ as
$X^*(B):=\inf_{B_{NC}\in NC_G} D(B||B_{NC})$, where $D$  denotes relative entropy of the boxes $B$ and $B_{NC}$ defined operationally via distinguishability of box $B$ from box $B_{NC}$ in \cite{Short-Wehener}. It would be interesting to relate such defined measure with $X_{max}$ and $X_{u}$. Note also, that following \cite{Joshi-broadcasting} it is easy to define and study notion of (anti)robustness of contextuality. This measure will be used in \cite{Pankaj-cont-broad}. It would be also interesting to investigate possible connection
between our measures and entropic tests of contextuality put forward in \cite{KurzynskiRK2012-entropic-context,ChavesFritz2012-entropic-context}
(which have their roots in entropic Bell inequalities \cite{BraunsteinC1988}).

Finally, we note that our measures can be useful for description of experimental results as they are based on correlations between measurement outcomes rather than on mutual exclusiveness of observables. It is important, since in practice it is very difficult to satisfy the latter condition in experiment.

\begin{acknowledgments}
We thank M.T. Quintino for helpful comments. A.G. and A.W. thank Pawe\l{}
Kurzy\'nski for useful discussion. This work is supported by ERC grant QOLAPS. KH also acknowledges grant BMN 538-5300-0975-12. PJ is supported by grant MPD/2009-3/4 from Foundation for Polish Science. Part of this work was done in National Quantum Information Centre of Gda{\'n}sk.
\end{acknowledgments}

\bibliographystyle{apsrev}
%\bibliography{rmp11-hugekey-phd}

\onecolumngrid %awesome option to make use of one column in revtex for some reason multicol package doesnt seem to work!!
\begin{appendix}
\section{Preliminaries}\label{sec:prelim}
We denote a hypergraph as $G:=(V_G,E_G)$ where $V_G=\{A_1,...,A_k\}$ is a set of $k$ observables and $E_G$ being a set of contexts of the hypergraph, i.e. the set of subsets of mutually commensurable observables of $V_G$. A box has an input ${\bf x}$ with cardinality $n$ equal to the number of edges of the hypergraph (number of contexts in a given $G$) and (for simplicity we assume) each output has the same cardinality $d$ of dimension equal to multiplication of cardinalities of outputs of $A_i$  which contribute in the corresponding context. The set of such boxes we denote as $B^{(k)}_G$. We say that a box is {\it compatible} with a hypergraph $G$ if it is family of $n$ probability distributions such that for each $c =\{A_{i_1},...,A_{i_{|c|}}\} \in E_G$, where $|c|$ is the power of the context $c$, there is a corresponding probability distribution in this family on  $\Omega(A_{i_1})\times...\times \Omega(A_{i_{|c|}})$. We denote it as a family of distributions $\{P({\bf a}|x_i)\}$ and $x_i\in E_G$.

{\definition For a given hypergraph $G=(V_G,E_G)$, $B \in B^{(k)}_G$ is a {\it consistent} box if for all pairs $c,c' \in E_G$, and for set of observables $S =  c \cap c' \neq \emptyset$ there is
\beq
\forall_s,\,\, \sum_{t} Pr(S=s,T=t|x=c) = \sum_{t'} Pr(S=s,T'=t'|x=c')
\eeq
where $T= c - S$ and $T' = c' - S$. The set of all consistent boxes compatible with hypergraph $G$ that has $n$ contexts is denoted as $C^{(n)}_G$.
\label{def:consistent}
}

Note, that the well known non-signaling condition is special case of such defined consistency.

{\definition A {\it non-contextual} box associated with a hypergraph $G$ is a consistent box with a property that there exists a common joint probability distribution for all the observables in $V_G$. The set of all such boxes compatible with $G$, we denote as
$NC_{G}$. All boxes that are consistent but do not satisfy this condition, we call
{\it contextual}.}

Similarly as in the main text, to specify distributions that belong to box $B \in C^{(n)}_G$ we will denote it as $\{g(\lambda_c)\}$ where $c$ numbers the contexts running from $1$ to $n$. If it is not stated otherwise, in what follows we assume $n\geq 3$, since for $n\leq 2$ all boxes compatible with any hypergraph $G$, are non-contextual.
If a box $B$ is non-contextual, we denote it as $\{p(\lambda_c)\}$, and by $p(\lambda)$ we will denote the joint probability distribution on $V_G$ (which exists by definition of non-contextual box) of which $p_c$'s are appropriate marginals. For short, by $p(\lambda) \in S$ for some set of boxes $S$ we mean that non-contextual box
defined by $p(\lambda)$ belongs to $S$ where graph $G$ with which this box is compatible should be understood from the context. We now make a trivial observation about these boxes:

{\observation A consistent box on $G=(\{A_1,...,A_k\},E_G)$ is non-contextual  iff it can be written as a convex combination of consistent deterministic boxes, i.e. such that the joint probability distribution of the outputs of all observables $A_1,...,A_k$ equals $\delta_{\bf a_0, a}$ for some fixed vector ${\bf a_0}$.}

%{\observation A noncontextual box on $G=(\{A_1,...,A_k\},E_G)$ is a mixture of boxes that are deterministic i.e. such that the joint probability distribution of the outputs of all observables $A_1,...,A_k$ equals $\delta_{\bf a_0, a}$ for some fixed vector ${\bf a_0$.}

{\it Proof}.

It follows from the definition of noncontextual boxes: the joint probability distribution of all observables $A_1,...,A_k$ is a mixture of the deterministic ones.\blacksquare

\section{Proof of equivalence}
\label{app:lem:prime}
In this section we present one of the main results of this work - equality of the mutual information of contextuality and the relative entropy of contextuality.
In this and the next section, for the sake of proof, we will use also a quantity defined on box $B =\{g(\lambda_c)\}$ as
$X_{\{p(c)\}}(B):= \min_{p(\lambda)}\sum_{c} p(c) D(g(\lambda_c) || p(\lambda_c))$, which is a version of relative entropy of contextuality for fixed distribution $\{p(c)\}$.

{\theorem For any box $B  =\{g(\lambda_c)\}\in C^{(n)}_G$, there holds $I_{max}(B) = X_{max}(B)$.
\label{thm:equivalence}
}

{\it Proof}.

To show the equality, we introduce  another measure of contextuality $I_{max}'$,
\beq
 I'_{max}(B) \equiv  \sup_{\{p(c)\}} \min_{\{g(\lambda|c):g(\lambda_c|c)=g(\lambda_c)\},p(\lambda)} \sum_{c} p(c) D(g(\lambda|c)||p(\lambda)),
\label{eq:equiv}
\eeq
and prove $X_{max}(B) = I'_{max}(B)= I_{max}(B)$.
The proof will not involve optimality of distribution $p(c)$ over which in all quantities we take supremum, so
we show the equality for $X_{\{p(c)\}}$, $I'_{\{p(c)\}} := \min_{\{g(\lambda|c):g(\lambda_c|c)=g(\lambda_c)\},p(\lambda)}
\sum_{c} p(c) D(g(\lambda|c)||p(\lambda))$ and $I_{\{p(c)\}}$, from which desired equality follows. We then fix $p(c)$ and $B\in C^{(n)}_G$ arbitrarily from now on,
and show that $X_{\{p(c)\}} = I'_{\{p(c)\}} = I_{\{p(c)\}}$. We prove now the first of these equalities. It is easy to see that
$I'_{\{p(c)\}}(B) \geq X_{{\{p(c)\}}}(B)$ since relative entropy does not increase under partial trace. To see the converse inequality,
consider the optimal classical probability in $X_{{\{p(c)\}}}$, call it $p^{*}(\lambda)$ (see \ref{app:not-equal} for the proof, that such $p^*(\lambda)$ exists) with marginals $p^{*}(\lambda_c)$, then
find a conditional probability distributions $p^{*}(\lambda_c'|\lambda_c)$ such that $p^{*}(\lambda_c'|\lambda_c)p^{*}(\lambda_c) = p^{*}(\lambda)$,
where $\lambda= \lambda_c'\lambda_c$, and define $g(\lambda|c) = p^{*}(\lambda_c'|\lambda_c)g(\lambda_c)$. It is easy to check,
that such a choice saturates the inequality $I'_{{\{p(c)\}}}(B) \geq X_{{\{p(c)\}}}(B)$ giving equality.

To see that $I_{{\{p(c)\}}}(B) = I'_{{\{p(c)\}}}(B)$,
we use the following fact:
\beq
I(\sum_c p(c) |c\>\<c|\ot {{\acal}_c}) \equiv \sum_c p(c) D( g(\lambda|c) || \sum_c p(c) g(\lambda|c)) = \min_{p(\lambda)} \sum_{c}p(c) D(g(\lambda|c)||p(\lambda)),
\label{eq:prime}
\eeq
where ${{\acal}_c}$ has distribution $g(\lambda|c)$, which is
proven in lemma \ref{lem:prime} below, stated in more general - quantum case (where in place of $g(\lambda|c)$ there is a quantum state $\rho_c$ and minimization is over some states $\sigma$). If we set minimization over $g(\lambda|c)$ having marginals $g(\lambda_c)$ of a box B, we get desired equality.

Summarizing the results we get $I_{\{p(c)\}}(B) = I'_{\{p(c)\}}(B) = X_{\{p(c)\}}(B)$ for arbitrary $p(c)$ and $B$, hence taking supremum over this distribution proves $I_{max}(B) = X_{max}(B)$ for arbitrary consistent box $B$.\blacksquare

Before proving equality (\ref{eq:prime}), we need another result, stated in the lemma below. We need it only for random variables, but we state it for quantum states, since it is valid for quantum states in general, and use the fact that quantum relative entropy and relative entropy distance coincide for classical distributions:

{\lemma For a quantum state $\rho$ with subsystems $Tr_B\rho \equiv \rho_A$ and $Tr_A\rho \equiv \rho_B$
\be
\inf_{\sigma_A,\sigma_B} S(\rho||\sigma_A\ot\sigma_B) = S(\rho||\rho_A\ot \rho_B)
\ee
where $Tr_A$ ($Tr_B$) denotes the partial trace over system $A$ ($B$), and $S$ is quantum relative entropy distance \cite{Nielsen-Chuang}.
\label{lem:mutual-prod}
}

{\it Proof.}

We first note, that $\log (\sigma_A\ot\sigma_B)) = (\log \sigma_A ) \ot I_B + I_A\ot (\log \sigma_B)$, where $I_A$ and $I_B$ are identity operators
on systems $A$ and $B$ respectively. Thus
\begin{multline}
S(\rho || \sigma_A\ot\sigma_B) = -S(\rho)  - Tr \rho \log (\sigma_A\ot\sigma_B) =\\ 
-S(\rho) + S(\rho_A) + [ - S(\rho_A) - Tr\rho_A\log\sigma_A] + S(\rho_B) + [ - S(\rho_B) - Tr\rho_B\log\sigma_B] = I(\rho) + S(\rho_A||\sigma_A) + S(\rho_B||\sigma_B)
\end{multline}
Where $I(\rho)$ is quantum mutual information \cite{Nielsen-Chuang}. The last equality proves that
$S(\rho || \sigma_A\ot\sigma_B) \geq I(\rho)$ because the relative entropy terms $S(\rho_A||\sigma_A)$ and $S(\rho_B||\sigma_B)$ are non-negative,
but $S(\rho||\rho_A\ot\rho_B) = I(\rho)$, hence the equality.\blacksquare

We prove now the lemma needed in proof of theorem \ref{thm:equivalence}.
We state it again for quantum states, since it is valid not only for probability distributions:

{ \lemma For arbitrary ensemble of quantum states $\{p(c),\rho_c\}$, there holds
\be
I(\sum_c p(c) |c\>\<c|\ot \rho_c) = \inf_{\sigma} \sum_c p(c) S(\rho_c||\sigma).
\ee
\label{lem:prime}
}

{\it Proof}.

Let us note that LHS can be rewritten as $S(\sum_c p(c) |c\>\<c|\ot \rho_c|| (\sum_c p(c)|c\>\<c|) \ot (\sum_c p(c) \rho_c))$.
Then, we use the fact that denoting $\sum_c p(c) |c\>\<c|\ot \rho_c$ as $\rho$, by lemma \ref{lem:mutual-prod} we have
\beq
S(\rho||(\sum_c p(c)|c\>\<c|) \ot (\sum_c p(c) \rho_c)) =\inf_{\sigma_A,\sigma_B} S(\rho||\sigma_A\ot\sigma_B).
\eeq
Knowing that $\sum_c p(c)|c\>\<c|$, i.e. the subsystem of $\rho$, is the best $\sigma_A$ in the above minimization, we can fix it, having
\ben
S(\rho||(\sum_c p(c)|c\>\<c|) \ot (\sum_c p(c) \rho_c)) = \inf_{\sigma} S(\rho||(\sum_c p(c)|c\>\<c|)\ot\sigma)
\een
It is then easy to check that the RHS of above equals just $\inf_{\sigma} \sum_c p(c) S(\rho_c||\sigma)$, and the assertion follows.\blacksquare

\section{Twirling and isotropic boxes. Simplifying computation of $X_u$}
\label{app:computing}

In order to compute $X_u$ for the isotropic xor-boxes %(see example (\ref{eq:fam})) 
and the KCBS box \cite{Klyachko-5}, we first observe that these boxes have numerous symmetries, i.e. they are invariant under some non-contextuality preserving operations. In this paragraph we specify groups of such operations and a map which applies them at random, called {\it twirling}. This leads us to the definition of isotropic boxes and the main result of this section (Theorem \ref{thm:iso-optimal}) which shows that for these boxes it is enough to minimize in the definition of $X_u$ only over non-contextual isotropic boxes.

To be more precise, consider any hypergraph $G$ with $n$ contexts and a box $B \in C^{(n)}_G$. A non-contextuality preserving operation satisfying $L(B) = B$ we call {\it non-contextuality preserving automorphism of $B$}. For any finite set of non-contextuality preserving automorphisms $\mathcal{L}$, if the group generated by the set $\mathcal{L}$ (denoted as $G_{\mathcal{L}}$) is finite of order $|G_{\mathcal{L}}|$, then the map defined on $B$ as
\be
B \stackrel{\tau^{\mathcal{L}}_B}{\longrightarrow} \sum_{l \in G_{\mathcal{L}}} { 1\over |G_{\mathcal{L}}| } l(B),
\ee
we call {\it B-$\mathcal{L}$-twirling} and denote as $\tau^{\mathcal{L}}_B$. The image of the set of all boxes through B-$\mathcal{L}$-twirling we call the set of {\it B-$\mathcal{L}$-isotropic states}:
\be
\mathcal{I}^{\mathcal{L}}_B := \{ D\in C^{(n)}_G: \exists_{F \in C^{(n)}_G} D = \tau^{\mathcal{L}}_B(F) \}.
\ee

Note, that there may be different twirlings depending on the set of generators $\mathcal{L}$ of $G_{\mathcal{L}}$. However, when the results are true for any fixed
choice of $\mathcal{L}$, or the set $\mathcal{L}$ is known from the context, we will omit it in notation, denoting the introduced objects as B-twirling ($\tau_B$), and a set of B-isotropic boxes ($\mathcal{I}_B$).

We observe that to find the set of B-isotropic boxes we need not to apply $\tau^{\mathcal{L}}_B$. By theorem \ref{thm:image}, which we prove below, the set $\mathcal{I}$ is equal to the set of boxes invariant under elements of $\mathcal{L}$. This theorem is true for any subset of linear space, but for clarity, we state it for the set of consistent boxes.

{\theorem For a hypergraph $G$ and the set of consistent boxes $C^{(n)}_G$ compatible with this graph, let $\mathcal{F}$ be a finite group  of linear maps $L:C^{(n)}_G\rightarrow C^{(n)}_G$ and $\mathcal{H} = \{h_1,...,h_n \} \subseteq \mathcal{F}$ a subset of its elements such that each of them have its inverse $h^{-1}_i$ in $\mathcal{F}$. Let us define a family of boxes $B$ invariant under transformations $h_i$:
\ben
\mathcal{T} := \{B \in C^{(n)}_G : \forall_i \, h_i (B) = B \},
\een
and a subgroup $\mathcal{F}_\mathcal{H} \subseteq \mathcal{F}$ generated by $\mathcal{H}$.
We then have the following:
\beq
\textrm{Im}_{\mathcal{F}_\mathcal{H}}(C^{(n)}_G) :=
\{D \in C^{(n)}_G: \exists_{B \in C^{(n)}_G} \,\sum_{f \in \mathcal{F}_\mathcal{H}}\frac{1}{|\mathcal{F}_\mathcal{H}|} f(B)  = D \} = \mathcal{T}.
\eeq
\label{thm:image}
}

\textit{Proof.}

Let $B \in \textrm{Im}_{\mathcal{F}_\mathcal{H}}(C^{(n)}_G)$. Then for each $i$ we have:
\ben
h_i(B) &=& h_i(\sum_{f \in \mathcal{F}_\mathcal{H}}\frac{1}{|\mathcal{F}_\mathcal{H}|} f(B) ) \\
       &=& \sum_{f \in \mathcal{F}_\mathcal{H}} \frac{1}{|\mathcal{F}_\mathcal{H}|} h_i\circ f(B)  \\
       &=& \sum_{\tilde{f}= h_i\circ f \in \mathcal{F}_\mathcal{H}} \frac{1}{|\mathcal{F}_\mathcal{H}|}\tilde{f}(B)  =B,
\een
where in first step we use linearity of the maps $h_i$ and in the last we use the fact that
$\tilde{f}$ runs through the whole group $\mathcal{F}_\mathcal{H}$ since each $h_i$ has its inverse. From the above we see that $B\in \mathcal{T}$, and so $\textrm{Im}_{\mathcal{F}_\mathcal{H}}(C^{(n)}_G) \subseteq \mathcal{T}$.

On the other hand, for each box $B \in \mathcal{T}$ we have:
\ben\label{eq:B}
B = \sum_{f \in \mathcal{F}_\mathcal{H}} \frac{1}{|\mathcal{F}_\mathcal{H}|} f(B) ,
\een
because for all $f \in \mathcal{F}_\mathcal{H}$, $f=h_{i_1} \circ ... \circ h_{i_n}$ ($h_{i_k} \in \mathcal{H}$), and so $f(B) = B$, from which we arrive at Eq.(\ref{eq:B}). Thus, we showed that $ \mathcal{T} \subseteq \textrm{Im}_{\mathcal{F}_\mathcal{H}}(C^{(n)}_G)$ which, jointly with the opposite inclusion, proves the theorem.\blacksquare

Consider now specific set of non-contextuality preserving automorphisms $\mathcal{L}_0$ which is any set of compositions of
 two types of linear maps: (i) $\pi_i$ - permutations of observables, and (ii) $b_i$
- negations of outputs of observables. For this set we have general theorem which allows for easier evaluating the relative entropy of contextuality.

{\theorem
For any box $B \in C^{(n)}_G$ and a set of B-$\mathcal{L}_0$-isotropic boxes $\mathcal{I}^{\mathcal{L}_0}_B$ we have:
\be\label{xu}
X_{u}(B)= \min_{p(\lambda)\in \mathcal{I}^{\mathcal{L}_0}_B} \sum_{c} {1\over n} D(g(\lambda_c) ||
p(\lambda_c)),
\ee
where the minimum is taken over all probability distributions $p(\lambda)$ which
give rise to non-contextual box from the set of B-$\mathcal{L}_0$-isotropic boxes $\mathcal{I}^{\mathcal{L}_0}_B$.
\label{thm:iso-optimal}
}

\textit{Proof.}

Let $p(\lambda)$ be optimal for $X_u(B)$, and denote the non-contextual box defined by this distribution as $B_{nc}$. Because of the choice of $\mathcal{L}_0$, for any element $f$ in group $G_{\mathcal{L}_0}$ generated by this set,
there is
\be
X_u (B) = \sum_{c} {1\over n}
D(g_f(\lambda_{c}) || p_f(\lambda_{c})),
\ee
where $g_f(\lambda_c)$ and $p_f(\lambda_c)$ are distributions of context $c$ of a box $f(B)$ and a box $f(B_{nc})$ respectively. To see this, we note that, by definition of $\mathcal{L}_0$, $f$ is a composition of permutation of observables and bit-flips of their outputs. It is then enough to prove separately that the above equality holds, for $f$ being one of them. Consider first $f$ to be a permutation of observables. Since $f(B) = B$, it is also an automorphism of $G$ with which $B$ is compatible, hence it is special permutation of observables which induces permutation of the contexts and in turn of elements $D(g(\lambda_c)|p(\lambda_c))$. It means that applying $f$ induces just change of the order of summation in the definition of $X_u$. Second, if $f$ is a bit-flip, since it is applied to
both $g(\lambda_{c})$ and $p(\lambda_{c})$, it does not change the relative entropy which is invariant under doubly applied
reversible operations \cite{Nielsen-Chuang}. Thus we have:

\beq
X_u (B) =  \sum_{c} {1\over n} \sum_{f\in \mathcal{F}_{\mathcal{L}_0}}
\frac{1}{|\mathcal{F}_{\mathcal{L}_0}|} D(g_f(\lambda_{c}) ||
p_f(\lambda_{c})) \geq
  \sum_{c} {1\over n}  D \left( \sum_{f\in \mathcal{F}_{\mathcal{L}_0}}
\frac{1}{|\mathcal{F}_{\mathcal{L}_0}|} g_f(\lambda_{c}) || \sum_{f\in \mathcal{F}_{\mathcal{L}_0}}
\frac{1}{|\mathcal{F}_{\mathcal{L}_0}|} p_f(\lambda_{c}) \right),
\label{xineq}
\eeq
where in the second line we used the joint convexity of relative entropy. What we
obtain is the fact that such process of symmetrization cannot increase the relative
entropy. What is more, since $f$ is an automorphism of $B$, we
have that for each context $c$:
\ben
\sum_{f \in \mathcal{F}_{\mathcal{L}_0}}\frac{1}{|\mathcal{F}_{\mathcal{L}_0}|} g_f(\lambda_{c}) = g(\lambda_c).
\een

We observe now, that since $\tau^{{\mathcal{L}}_0}_B$ preserves non-contextuality,  the box $\tau^{{\mathcal{L}}_0}_B(B_{nc})$ has a context $c$ equal to
$\sum_{f\in \mathcal{F}_{\mathcal{L}_0}}\frac{1}{|\mathcal{F}_{\mathcal{L}_0}|} p_f(\lambda_{c})$ and is a
non-contextual box. Since $B_{nc}$ is optimal for $X_u$
and, when we substitute the box $\tau^{{\mathcal{L}}_0}_B(B_{nc})$ in place of $B_{nc}$, we cannot increase the quantity $X_u$ due to inequality (\ref{xineq}),
the box $\tau^{{\mathcal{L}}_0}_B(B_{nc})$ must also be optimal for $X_u$, which proves desired equality in (\ref{xineq}). We have $\tau^{{\mathcal{L}}_0}_B(B_{nc})\in \mathcal{I}^{\mathcal{L}_0}_B$ hence the assertion follows.\blacksquare

\section{Computing $X_u$ for the exemplary isotropic xor-boxes}
\label{app:beta}

In this section we specify twirling operations $\mathcal{L}_0$ for the xor-boxes, hence showing that one can obtain the isotropic xor-boxes 
%introduced in (\ref{eq:fam}) : Here may be we could say  "introduced in the main paper" perhaps!!
by operations that are non-contextuality preserving. This is crucial, since then we can use Theorem \ref{thm:iso-optimal} to compute $X_u$ for these boxes, which is done in Theorem \ref{thm:iso-xu-values}. We first define the Peres-Mermin's (PM) and Mermin's (M) box below:

The $PM$ is a box on
$ G_{PM} = (\{A_{1},...,A_{9}\},\{\{A_1,A_2,A_3\},\{A_4,A_5,A_6\}, \\
\{A_7,A_8,A_9\}, \{A_1,A_4,A_7\},\{A_2,A_5,A_8\},\{A_3,A_6,A_9\}\})
$
with $g(\lambda_c) = P^{(3)}_{even}$ for first 5 contexts, and  $g(\lambda_c) = P^{(3)}_{odd}$ for the 6th one \cite{Peres1990-KS,Mermin1990-KS}.

The $M$ is a box on
$
G_{M} = (\{A,B,C,D,E,a,b,c,d,e\}, \{\{B,e,a,D\},\{D,b,c,A\}, \\
\{A,d,e,C\},\{C,a,b,E\},\{E,c,d,B\}\})
$
with $g(\lambda_c) = P^{(4)}_{even}$ for first 4 contexts, and $g(\lambda_c) = P^{(4)}_{odd}$ for the 5th one \cite{Mermin-rev}.

\begin{figure}%[!hbt]
\begin{center}
\subfloat[PM box]{\label{figurynka:pm}\includegraphics[width=0.2\textwidth]{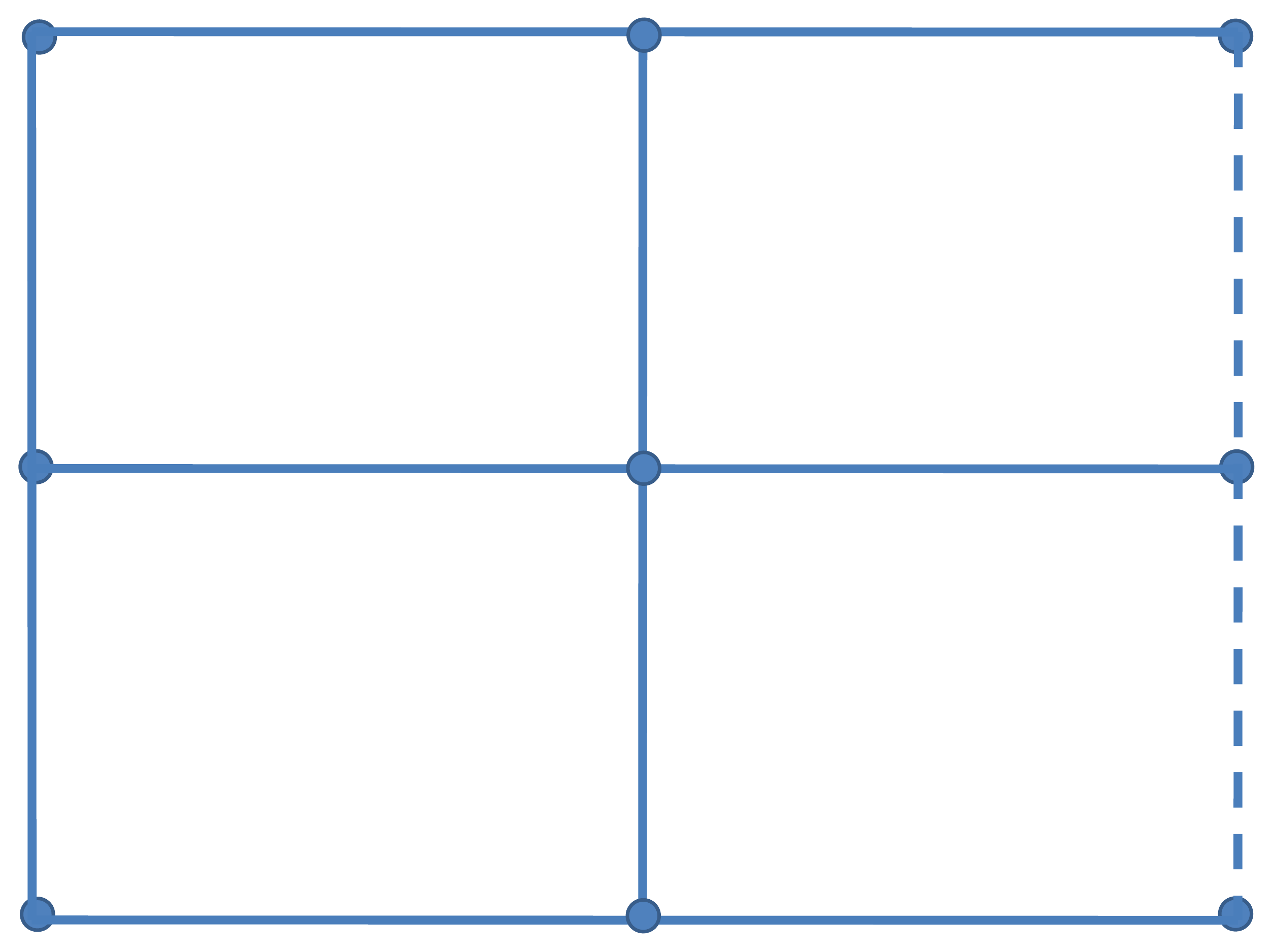}}
\qquad
\subfloat[M box]
{\label{figurynka:m}\includegraphics[width=0.2\textwidth]{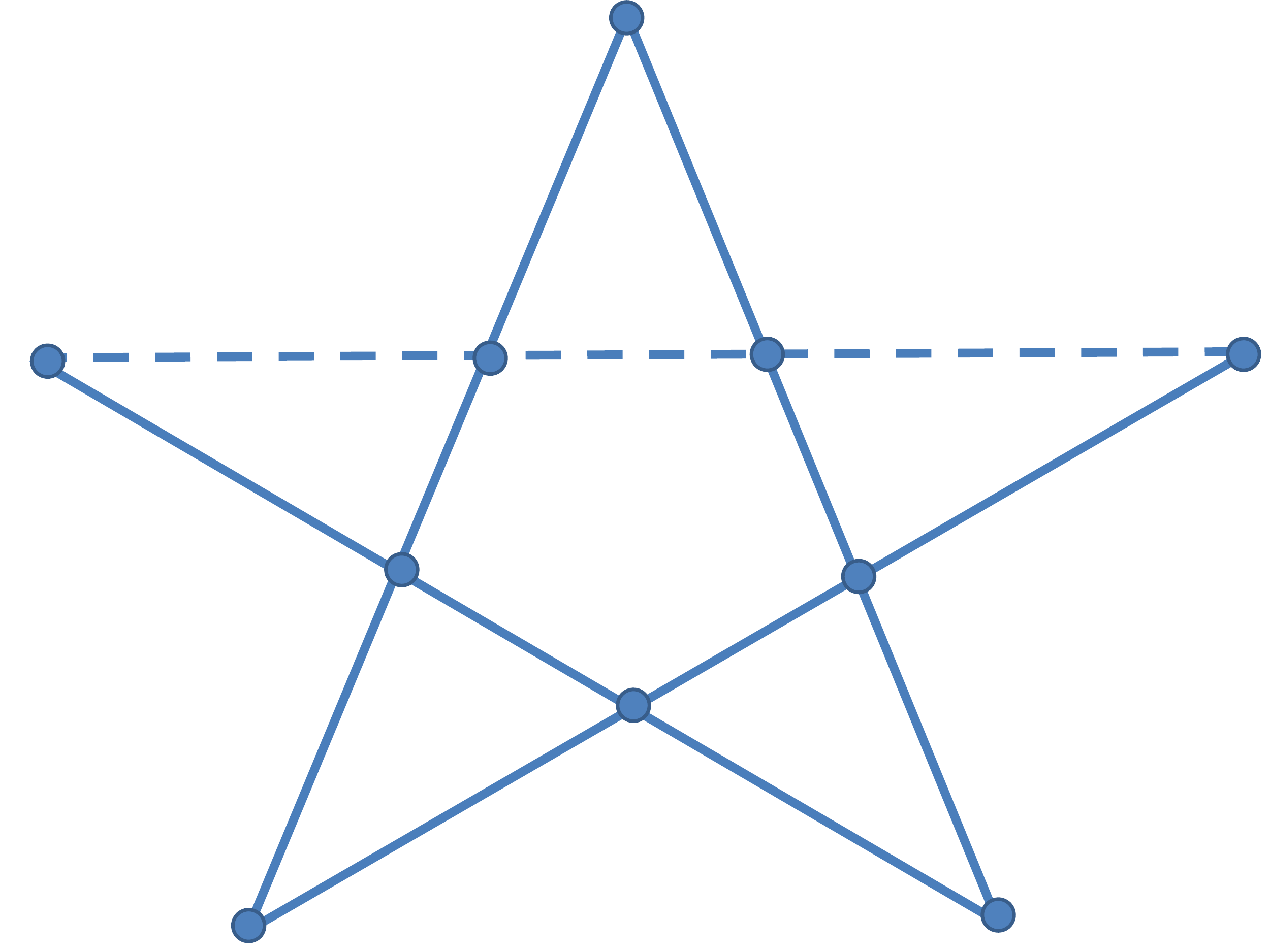}}
\end{center}
\caption{{\small
Depiction of the hypergraphs of the Peres-Mermin box (a) and Mermin box (b). Vertices denotes observables. Each solid line corresponds to a context with fully correlated distribution, dashed one with fully anti-correlated distribution.
}}
\label{pm}%do remember that i have changed this label to pm but didnt find where it has been used because then i need to change label to this new one
\end{figure}

To begin with, we introduce twirling for $PM$ box by specifying the set $\mathcal{L}_0$ which leads to one-parameter family of isotropic boxes.
\begin{itemize}
\item
For PM box we choose $\mathcal{L}_0 = \{ h_1, ..., h_8 \}$, where the elements $h_i$ are in general compositions of maps (i) and (ii): $h_1,...,h_6$ - are 3! permutations of contexts $\{A_1,A_2,A_3\}, \{A_4,A_5,A_6\}, \{A_7,A_8,A_9\}$ i.e. the rows on Fig. \ref{figurynka:pm}, $h_7$ - a swap of  $\{A_1,A_4,A_7\}$ and $\{A_2,A_5,A_8\}$, i.e. a swap of two solid columns on Fig. \ref{figurynka:pm}, $h_8$ - is a composition of permutation defined by mappings: $A_4 \leftrightarrow A_2$, $A_7 \leftrightarrow A_3$, $A_7 \leftrightarrow A_6$ (the rest of the observables are mapped to themselves), composed with a bit-flip of the output of observable $A_9$. This operation is a reflection of the hypergraph w.r.t. to the diagonal with appropriate bit-flip. on Fig. \ref{figurynka:pm}. The set
$\mathcal{I}^{\mathcal{L}_0}_{PM}$ we call the set of isotropic $PM$ boxes. The reason for this is stated in lemma below:
\end{itemize}

{\lemma There holds:
\be
\mathcal{I}^{\mathcal{L}_0}_{PM}= \{\alpha PM + (1-\alpha) PM'| \alpha \in [0,1]\},
\ee
where $PM'$ is an opposite version of the box $PM$, i.e. $PM$ with $P^{(3)}_{even}$ exchanged with $P^{(3)}_{odd}$ and vice versa.
\label{lem:iso:equal-pm}
}

{\it Proof}.

To see the above statement, we will use Theorem \ref{thm:image}. Due to this theorem it is enough to argue that invariance of a box under $\mathcal{L}_0$ implies that it belongs to $\mathcal{I}^{\mathcal{L}_0}_{PM}$. In the proof we will refer to Fig. \ref{figurynka:pm}. First, due to invariance under $h_1,...,h_6$ the rows need to have the same probability distribution. Second, using $h_8$ we obtain that middle row and middle column has the same distributions. Third, by $h_7$ we get that all solid lines has the same distributions with 8 probabilities $q(ijk)$ of string $(ijk)$ where $i,j,k$ are binary. Due to invariance under $h_1,...,h_6$, both the solid columns and the dashed column are permutationally symmetric, i.e. are described only by $q(000)$, $q(001)$, $q(011)$ and $q(111)$ ($r(000)$, $r(001)$, $r(011)$ and $r(111)$ for dashed column). Invariance under operation $h_8$ imposes $q(000)=r(001)$ and $q(011)=r(010)$ which equalizes $q(000)$ and $q(011)$ because of $r(001)=r(010)$. Thus $q(000) = \alpha/4$ for some parameter $\alpha \in [0,1]$.
Similarly, we have $q(111)=r(110)$, $q(010)=r(011)$,
which implies that $q(111)=q(010)\equiv (1-\alpha)/4$. Exchanging $p$ with $q$, we get
that also $r(000)=r(011)\equiv(1-\alpha)/4$ and $r(111)=r(011)\equiv \alpha/4$, which ends the proof.\blacksquare

The argument given in the above lemma is analogous in the case of other xor-boxes considered in this paper, where in particular we have:
\begin{itemize}
\item
For $M$ box we choose $\mathcal{L}_0 = \{ \tilde{h}_1,...,\tilde{h}_{10} \}$, where:
$\tilde{h}_1$ - reflection of the star with respect to the $Aa$ symmetry line, $\tilde{h}_2$ - reflection of the star with respect to the $Cc$ symmetry line with bit flip on the node $c$, $\tilde{h}_3$ - reflection of the star with respect to the $Dd$ symmetry line with bit flip on the node $d$, $\tilde{h}_4$ - reflection of the star with respect to the $Ee$ symmetry line with bit flip on the node $E$, $\tilde{h}_5$ - reflection of the star with respect to the $Bb$ symmetry line with bit flip on the node $B$, $\tilde{h}_{6-10}$ - bit flips on three nodes that form any triangle on the hypergraph ($Acd$, $Bde$, etc.). For such defined $\mathcal{L}_0$ there holds:
\be
\mathcal{I}^{\mathcal{L}_0}_{M} = \{\alpha M + (1-\alpha) M' | \alpha \in [0,1] \},
\label{lem:iso:equal-m}
\ee
and the set of these boxes we call isotropic $M$ boxes.
\item
For $CH_{(n)}$ box we choose $\mathcal{L}_0 = \{ \hat{h}_1,...,\hat{h}_j,..., \hat{h}_{n-1} \}$, where: $\hat{h}_j$ is a composition of cyclic permutation of contexts such that all $\{A_{i}, A_{i+1} \} \longrightarrow \{A_{i+j}, A_{i+j+1} \}$ with bit flips on the observables $A_1,...,A_j$. For such defined $\mathcal{L}_0$ there holds:
\be
\mathcal{I}^{\mathcal{L}_0}_{CH^{(n)}}=\{\alpha CH_{(n)} + (1-\alpha) CH_{(n)}'| \alpha \in [0,1]\},
\label{lem:iso:equal-ch}
\ee
and the set of these boxes we call isotropic $CH_{(n)}$ boxes.
\end{itemize}

Let us now fix a contextual box $B = \{g(\lambda_c)\}$ and denote by $g(\lambda_c)_i$ the outcome $i$ of distribution $g(\lambda_c)$ under a measurement on the box $B$ the context $c$. For a box $\tilde{B} = \{\tilde{g}(\lambda_c)\}$ compatible with the same hypergraph $G$ as $B$, we define the quantity $\beta_B$ which measures how contextual is box $\tilde{B}$ w.r.t. to box $B$:
\ben
\beta_{B}(\tilde{B}) := \sum_c \sum_{i \in \textrm{supp}( g(\lambda_c))} \tilde{g}(\lambda_c)_i  ,
\een
where $g(\lambda_c)$ are probabilities of outcomes within a given context, and $supp(g(\lambda_c))$ is the support of the distribution $g(\lambda_c)$, i.e. the set of the outcomes of a measurement of the context $c$ which have nonzero probability in distribution $g(\lambda_c)$.

We will need some properties of $\beta_B(.)$, which are collected in the lemma below, where we treat boxes as vectors of probabilities.
{\lemma
For any box $\tilde{B} \in C^{(n)}_G$ and an xor-box $B \in C^{(n)}_G$ with all contexts of the same cardinality $m$ and all observables of the same cardinality $2$, there holds:
\be
\beta_B(\tilde{B}) = 2^{(m-1)}\<\tilde{B}|B\>,
\ee
where $\<.|.\>$ is Euclidean scalar product of vectors.
Moreover, for a twirling $\tau_B^{\mathcal{L}_0}$ there holds:
\be
\beta_B(\tilde{B}) = \beta_B(\tau_B^{\mathcal{L}_0}(\tilde{B})).
\ee
\label{lem:bet-prop}
}

{\it Proof}.

The first statement is easy, as $2^{(m-1)}B$ is a vector of $1$'s for probabilities for which $g(\lambda_c) >0$ where $B=\{g(\lambda_c)\}$, hence the scalar product
sums the probabilities of box $\tilde{B}$ from the support of box $B$. To see the next consider
the following chain of equalities:
\begin{multline}
\beta_B(\tau_B^{\mathcal{L}_0}(\tilde{B})) = 2^{(m-1)}\< \sum_{f\in \mathcal{F}_{{\mathcal L}_0}} {1\over |\mathcal{F}_{{\mathcal L}_0}|} f(\tilde{B}) |B\> 
= \sum_{f\in \mathcal{F}_{{\mathcal L}_0}}{1\over |\mathcal{F}_{{\mathcal L}_0}|} 2^{(m-1)} \< f(\tilde{B})|B\> 
= \sum_{f\in \mathcal{F}_{{\mathcal L}_0}}{1\over |\mathcal{F}_{{\mathcal L}_0}|} 2^{(m-1)} \< f(\tilde{B})|f(B)\>  \\
= \sum_{f\in \mathcal{F}_{{\mathcal L}_0}}{1\over |\mathcal{F}_{{\mathcal L}_0}|} 2^{(m-1)} \< \tilde{B}|B\> = \beta_B(\tilde{B})
\end{multline}
where in the second equality we use linearity of scalar product, in the third we use the fact that by definition of twirling $f$ is an automorphism of $B$ and in the fourth we use the fact, that each $f$ is a composition of elements from ${\mathcal L}_0$, i.e. permutations of observables and bit flips of outputs,
hence it is a permutation, which does not change the scalar product.\blacksquare

Based on $\beta_B$ we can build naturally a contextuality inequality, which for $PM$ box is equivalent to that given in \cite{Cabello-first-ineq}, for PR box that given in \cite{CHSH} and for $CH_{(n)}$ box that from \cite{Aruajo-chain} (see also \cite{BraunsteinC1988}).

{\theorem
For an xor-box $B \in C^{(n)}_G$ with a single context with distribution $P_{odd}$, such that each vertex from $V_G$ belongs to even number of contexts
and for a non-contextual box $\tilde{B} \in C^{(n)}_G$, there holds:
\ben
\beta_B(\tilde{B}) \leq n-1,
\label{eq:beta-n-1}
\een
and the bound is tight.
\label{thm:beta-n-1}
}

\textit{Proof.}

In what follows, we generalize the argument of N.D. Mermin \cite{Mermin-rev}, with the use of which He proved that $M$ box is contextual.
Since any noncontextual box is a mixture of deterministic boxes, and by lemma \ref{lem:bet-prop}, $\beta_B(\tilde{B})$ is linear, it suffices to prove the above inequality for deterministic ones. Surely, deterministic boxes can attain only discrete values of LHS. Suppose then, that for noncontextual box $\textrm{LHS}=n$, i.e. all constraints of a contextual box are satisfied, meaning that for $n-1$ contexts the sum of outputs $\bigoplus_i a_i = 0$ ($even$) and for $1$ context $\bigoplus_i a_i = 1$ ($odd$), which gives a total sum over all contexts 1. On the other hand, for deterministic assignment, summing all the values for the whole hypergraph we get $\bigoplus_i n_i a_i = 0$ since each $n_i$ (the number of contexts to which the observable $A_i$ belongs to) is an even number by the assumption. This gives desired contradiction. The value of $RHS$ can be attained deterministically, e.g. by putting all the outcomes equal 0, which simultaneously tighten the inequality.\blacksquare

{\theorem
For an xor-box $B \in C^{(n)}_G$ with even $n$ and a simple context with distribution $P_{odd}$, such that each vertex from $V_G$ belongs to even number of contexts and for a non-contextual $\tilde{B} \in C^{(n)}_G$, we have:
\ben
\beta_B(\tilde{B})\geq 1.
\een
\label{thm:beta-1}
Moreover if the number of vertices in each context is odd then the bound is tight.
}

\textit{Proof.}

The argument is analogous to the proof of Theorem \ref{thm:beta-n-1}. Again, we only need to consider deterministic assignments. Suppose there is a deterministic assignment of outcomes with $\textrm{LHS}=0$. Then the box would satisfy all the constraints of contextual opposite version of a box B. For this box, $n-1$ contexts has the sum of outputs equal to $\bigoplus_i a_i = 1$ ($odd$) and for $1$ context $\bigoplus_i a_i = 0$ ($even$), which gives a total sum over all contexts 1. This however is in contradiction with the fact that the sum over all vertices is 0 since each vertex appears an even number of times in the sum. Hence $\textrm{LHS} \geq 1$. To see the tightness in a special case, we observe that setting each vertex value 1 constitutes a deterministic assignment that has $\beta_B$ equal to 1. Indeed, since each context has an odd number of vertices, each edge has distribution $P_{odd}$, and exactly one of them is in accordance with box $B$.\blacksquare

We note, that the assumption about evenness of $n$ in the above theorem is necessary:

{\observation For $B \in \{M, CH_{(n)}\}$ with odd $n$, there exists a non-contextual box $\tilde{B}$ such that $\beta_B(\tilde{B}) = 0$.
\label{obs:zero}
}

\textit{Proof.}

There exists a deterministic assignment which sets $\beta_B$ to zero: first, we set all observables to 1, and then change into 0 $k$ of those which does not belong to context which has $P_{odd}$ in B, but such that each belong to disjoint pair of contexts. Such an assignment creates an opposite version of a box $B$, hence giving $\beta_B(\tilde{B}) =0$.\blacksquare

As we have seen, all examples of sets of isotropic xor-boxes considered so far are one parameter. We now prove the lemma which bounds this parameter for non-contextual boxes.

{\lemma For a non-contextual box $B_{\alpha} \in \mathcal{I} \equiv I^{\mathcal{L}_0}_{PM} \cup I^{\mathcal{L}_0}_{M}\cup I^{\mathcal{L}_0}_{CH_{(n)}}$ with $n$ contexts, there holds:
\be
\alpha \leq {n -1 \over n}.
\label{eq:r-bound}
\ee
For even $n$ there holds additionally:
\be
\alpha \geq {1\over n},
\label{eq:l-bound}
\ee
while for odd $n$ there is
\be
\alpha \geq 0,
\label{eq:z-bound}
\ee
and the bounds are tight.
\label{lem:bounds}
}

{\it Proof.}

By the definition of $\beta_B$, for any xor-box $B$ we have $\beta_{B}(B') = 0$ where $B'$ is an opposite version of box $B$ (with $P^{(m)}_{even}$ in place of $P^{(m)}_{odd}$ and vice versa). This implies that for any isotropic xor-box $B_\alpha \in\mathcal{I}$ we have
 \be
 \beta_B(B_\alpha) = n \alpha,
 \label{eq:alpha-beta}
 \ee
and in particular, for non-contextual isotropic boxes $B_\alpha \in\mathcal{I}$ by Theorem \ref{thm:beta-n-1} we have
 \be
 \alpha \leq {n-1\over n}.
 \label{eq:upper-alpha}
 \ee
To prove the second inequality, we observe, that $PM$ and $CH_{(n)}$ for even $n$ satisfies the assumption of Theorem \ref{thm:beta-1}, which gives the inequality (\ref{eq:l-bound}) in analogous way. The last inequality follows from the dependence (\ref{eq:alpha-beta}) and observation \ref{obs:zero}. To see that the boundary values of $\alpha$ are attained by non-contextual boxes, we first observe that
by theorem \ref{thm:beta-n-1}, there exists a non-contextual box $\tilde{B}$ with $\beta_B(\tilde{B})= n-1$.
Now by lemma \ref{lem:iso:equal-pm} equalities (\ref{lem:iso:equal-m}) and (\ref{lem:iso:equal-ch}) after twirling $\tau^{\mathcal{L}_0}_B$ $\tilde{B}$ belongs to ${\mathcal I}^{\mathcal{L}_0}_B$, hence it has a form $\alpha B + (1-\alpha) B'$ where $B'$ is opposite version of B. By lemma \ref{lem:bet-prop}, $\beta_B(\tilde{B}) = \beta_B(\tau^{\mathcal{L}_0}_B(\tilde{B}))=n-1$. Now, by equation (\ref{eq:alpha-beta}), we have $ \beta_B(\tau^{\mathcal{L}_0}_B(\tilde{B}))=n\alpha$ proving that $\tau^{\mathcal{L}_0}_B(\tilde{B})$ attains the value ${n-1}\over n$ of $\alpha$. This box is clearly non-contextual, since $\tau^{\mathcal{L}_0}_B$ is application at random some permutation of observables composed with bit-flips on outputs of observables, hence preserving non-contextuality. Analogous argument, by use of theorem \ref{thm:beta-1} and observation \ref{obs:zero}, proves the tightness of the bounds (\ref{eq:l-bound}) and (\ref{eq:alpha-beta}) respectively.
\blacksquare

%%%%%%%%%%%%%%%%%%%%%%%%%%%%% we cut out here the lemma about single-term

%%%%%%%%%%%%%%%%%%%%%%%%%%%%%%%%%%%%%%%%%%%%%%%%%%%%%% end of cut out the lemma about single-term

We can state the main theorem of this section:

{\theorem For $B_{\alpha} \in PM_{\alpha} \cup CH^{(n)}_{\alpha} \cup M_{\alpha}$ with $n\geq 3$ number of contexts and $\alpha \geq {n-1\over n}$ there holds
\be
X_{u}(B_{\alpha}) =
\log [(n-1)^{-\alpha}n] -h(\alpha),
\label{eq:xu-n-1}
\ee
while for even $n$ and $\alpha \leq {1\over n}$ there holds
\be
X_{u}(B_{\alpha}) =
\log [(n-1)^{(\alpha -1)}n] - h(\alpha),
\label{eq:xu-1}
\ee
where $h(\alpha) = -\alpha \log\alpha - (1-\alpha)\log(1-\alpha)$.
\label{thm:iso-xu-values}
}

{\it Proof.}

We first note, that in both cases we want to consider, the isotropic boxes satisfy assumptions of Theorem \ref{thm:iso-optimal}, hence
we need to optimize only over appropriate isotropic xor-boxes:
\be
X_{u}(B_{\alpha})= \min_{p(\lambda)\in \mathcal{I}^{\mathcal{L}_0}_{B_{\alpha}}} \sum_{c} {1\over n} D(g(\lambda_c) || p(\lambda_c)),
\label{xu2}
\ee
where for short by $p(\lambda)\in \mathcal{I}^{\mathcal{L}_0}_{B_{\alpha}}$ we mean some isotropic non-contextual box $B_{\alpha_0}$ which is defined by distribution $p(\lambda)$.

Since $B_{\alpha_0}$ has only two kinds of distributions $p(\lambda_c)$, $P^{\alpha_0}_{even}\equiv \alpha_0 P_{even} + (1-\alpha_0)P_{odd}$ and
$P^{\alpha_0}_{odd}\equiv \alpha_0 P_{odd} + (1-\alpha_0)P_{even}$ we can write:
\beq
 X_{u}(B_{\alpha}) = \min_{\alpha_0} \frac{1}{n}
 ( (n-1) D(P^{\alpha}_{even} || P^{\alpha_0}_{even}  )
 +    D(P^{\alpha}_{odd} || P^{\alpha_0}_{odd}  ) ),
\eeq
where $\alpha_0$ is bounded such that $B_{\alpha_0}$ is noncontextual, and $P^{\alpha}_{even}$ and $P^{\alpha}_{odd}$ are defined analogously to $P^{\alpha_0}_{even}$ and $P^{\alpha_0}_{odd}$, respectively.
Now, it is easy to check that if $S = \mathcal{I}^{\mathcal{L}_0}_{B_{\alpha}}$ for $B_{\alpha} \in PM_{\alpha} \cup M_{\alpha} \cup CH^{(n)}_{\alpha}$, the assumption of the lemma \ref{lem:one-term} are satisfied with $\Pi_c$ being identity operations for every $c$ for $n-1$ contexts for which $B_{\alpha}$ has the same distribution and bit-flip on one of the observables on a distribution of the remaining context, giving:
\be
 X_{u}(B_{\alpha}) =
 \min_{\alpha_0}  D(P^{\alpha}_{even}|| P^{\alpha_0}_{even}),
 \label{eq:xu-one-term}
\ee
where $\alpha_0$ is bounded by the fact that $P^{\alpha_0}_{even}$ is a distribution of a non-contextual box which is isotropic. More specifically, it is bounded according to lemma \ref{lem:bounds}, i.e. we have $ \alpha_0 \leq {n-1\over n}$. It is easy to check that for $\alpha \geq \alpha_0$ the function (\ref{eq:xu-one-term}) is decreasing with $\alpha_0$. Lemma \ref{lem:bounds} shows that the boundary value $\alpha_0 = {(n-1)\over n}$ is attained by non-contextual isotropic xor-box, hence the function attains minimum for this value of $\alpha_0$, which proves (\ref{eq:xu-n-1}). For $\alpha \leq \alpha_0$, this function is increasing, and again by lemma \ref{lem:bounds} attains minimal value at $\alpha_0 = {1\over n}$, which proves (\ref{eq:xu-1}).
\blacksquare

We note here, that due to the above theorem, $X_u(B_{\alpha})$ for even $n$, $\alpha \geq {(n-1)\over n}$ equals the value of $X_u(B_{\alpha'})$ for $\alpha' = 1-\alpha$, in correspondence with the fact that $B_{\alpha}$ can be changed by bit-flips into $B_{1-\alpha}$ which does not change the relative entropy distance.

In particular, for considered examples of xor-boxes i.e. in the case $\alpha = 1$ we have:
\ben
X_{u}(PR) = \log \frac{4}{3} \approx 0.4150, \\
X_{u}(PM) = \log \frac{6}{5} \approx 0.2630, \\
X_{u}(M) = \log \frac{5}{4} \approx 0.3219, \\
X_{u}(CH_{(n)}) = \log \frac{n}{n-1}.
\een
According to the above formula, $X_u$ tends to zero in asymptotic limit for maximally contextual chain boxes. Interestingly, if we do not take average over number of contexts, i.e. consider a measure $X(B):= \sum_c D(g(\lambda_c)||p(\lambda_c))$, it will equal to $n\log (1 + {1\over n-1})$ and tend asymptotically to $\log_2 e$ where $e$ is the Euler number. In other words, if we consider natural logarithm in definition of relative entropy, $X$ tends to 1 with increasing $n$. It means that, although "average" contextuality of chain box - per number of contexts - vanishes with increasing $n$, the "total" contextuality is bounded by 1 from below.
Remarkably, the same result holds for quantum maximally contextual chain boxes: $X(CH^{(n)}_{\alpha})$ based on natural logarithm tends to 1 for both odd and even $n$ where $\alpha$ for each $n$ is given in description of the \ma Fig \ref{fig:miary_CHn}.

For comparison, in the case of maximal violation of CHSH inequality we have $B_{CHSH} \equiv CH^{(4)}_\alpha$ with $\alpha=\cos^2 \frac{\pi}{8}$ which gives:
\ben
X_{u}(B_{CHSH}) \approx 0.0463.
\een

\section{Direct sum of boxes $X_u$ and $X_{max}$ are equal for isotropic xor-boxes but not equal in general}.
\label{app:not-equal}

As it was mentioned, $X_{max} \ge X_u$. In this section we prove that these two measures are equal for isotropic xor-boxes but are different in general . More precisely, we find their values on {\it direct sum} of two boxes, in terms of their values of the boxes themselves in theorem \ref{thm:direct-sum}.  Using this result, we show among others that these measures differ on any direct sum of contextual box and non-contextual one in corollary \ref{cor:xu-c-nc-differ}. To begin with, we show that in \ma equation \ref{eq:xu} we can indeed write minimum instead of infimum. We will show more, proving that in definition of $X_{\{p(c)\}}$ one can also
consider minimum. Taking than uniform distribution, we get the thesis for $X_{u}$. We also prove that both measures are faithful and are monotonous under certain subclass of noncontextuality preserving operations.

Recall that for a box $B=\{g(\lambda_c)\}$,
\be
X_{\{p(c)\}}(B)= \min_{\{p(\lambda)\}} \sum_{c \in E_G} p(c) D(g(\lambda_c) || p(\lambda_c))
\ee
which depends on both the box $B$ and probability distribution over contexts $\{p_c\}_c$.
Recall here also, that the minimization is taken over all joint probability distributions $p(\lambda)$
defined on outputs of observables, and $p(\lambda_c)$ are marginals, restricted to the context $c$.
This quantity can be written as  (using quantum notation for classical distributions)
\be
X_{\{p(c)\}}(B)=\inf_\sigma S(\rho|\sigma)
\ee
where $\rho=\sum_c p_c |c\>\<c| \ot \rho_c$ and $\sigma=\sum_c p_c |c\>\<c| \ot \rho_c$
with $\sigma_c$ representing probability distributions $p(\lambda_c)$, and $\rho_c$
 - the distributions $g(\lambda_c)$ of the box $B$.
One finds that the set of states $\sigma$ is convex and compact (note, that $\{p_c\}$
is fixed here). Therefore, since relative entropy is lower semicontinuous \cite{er-lower-cont}, there exists  state $\sigma^*$,
which achieves the infimum. This finishes the proof.

\subsection{Faithfulness and partial monotonicity of $X_u$ and $X_{max}$}
\label{app:props-of-measures}

We first argue, that both $X_u$ and $X_{max}$ are {\it faithful} i.e. that are nonzero iff the box is contextual.
Indeed, the relative entropy is lower bounded by square of variational distance between the probability distributions, which is zero only if all $g(\lambda_c)$ are equal to $p(\lambda_c)$. This however cannot hold for there is no single joint probability distribution with marginals $g(\lambda_c)$.

We now prove monotonicity of $X_u$ and $X_{max}$ under operations which lie within the set of non-contextuality preserving ones.
Given a box $B= \{g(\lambda_c)\} \in C^{(n)}_G$,
we define an operation $\Lambda$ as a mixture with some probabilities $p_j$ of independent channels $\Lambda_i^{(j)}$ acting on observable $i \in \{1,...,n\}$.
It is easy to see, that such defined set of operations is a convex subset of non-contextuality preserving operations.
To see that $X_{u}$ and $X_{max}$ are monotonous under these operations we need to use the fact that both measures can be expressed in terms of mutual information (\ma equation \ref{eq:mutual-form}). Application of $\Lambda$ is then a local action on variable ${{\acal}_c}$, and monotonicity of both $X_{u}$ and $X_{max}$
follows directly from data processing inequality \cite{CoverThomas}.

\subsection{ $X_{u}$ and $X_{max}$ are equal for isotropic boxes }

We will now need another fact that apart from Theorem \ref{thm:iso-optimal}, simplifies computation of $X_u$.

{\lemma
Let $B = \{g(\lambda_c)\} \in C^{(n)}_G$ and
\be
X_u(B) =
\inf_{p(\lambda) \in S} \sum_c {1\over n} D(g(\lambda_{c})| p(\lambda_{c})),
\ee
for some $S \subseteq NC_G$. If for any $p(\lambda) \in S$ there exist reversible operations $\Pi_c$ satisfying $\Pi_c(p(\lambda_c))= p(\lambda_{c_0})$ for all contexts $c$ and simultaneously $\Pi_c(g(\lambda_c))= g(\lambda_{c_0})$ for all contexts $c$, then
\be
X_u(B) =
\inf_{p(\lambda) \in S} D(g(\lambda_{c_0})| p(\lambda_{c_0})).
\ee
\label{lem:one-term}
}

{\it Proof.}

The proof boils down to the observation that relative entropy is invariant under bilateral reversible operations \cite{Petz-equality}.\blacksquare

We observe, that for all isotropic xor-boxes introduced in previous section, the assumption of the above lemma are satisfied. This enables us to prove the following important result:

{\theorem For isotropic boxes $B \in {\cal I}^{{\cal L}_0}_M\cup {\cal I}^{{\cal L}_0}_{PM}\cup {\cal I}^{{\cal L}_0}_{CH^{(n)}}$, there is $X_{max}(B)= X_u(B)$.
\label{thm:xmax-xu}
}

{\it Proof.}

Let us calculate the quantity $X^*(B):=\sum_c p(c) D( g(\lambda_c)||p(\lambda_c) )$ defined for a given $p^*(\lambda)$, which is optimal for the measure $X_u(B)$. According to the assumptions of Lemma \ref{lem:one-term} valid for isotropic boxes, we can write
\ben
X^*(B) &=& \sum_c p(c) D( g(\lambda_{c_0})||p(\lambda_{c_0}) )  \\
        &=& D( g(\lambda_{c_0})||p(\lambda_{c_0}) ) = X_u(B). \nonumber
\een

On the other hand, from the definition of $X_{ \{p(c) \} }$ we have
\ben
X_{ \{p(c) \} }(B) \leq X^*(B),
\een
hence
\ben
X_{ \{p(c) \} }(B) \leq X_u(B),
\een
and taking the supremum gives $X_{ max }(B) \leq X_u(B)$ while combining this with the fact $X_{ max }(B) \geq X_u(B)$ we obtain the equality of the two measures for isotropic boxes. \blacksquare

\subsection{$X_{max}$ and $X_{u}$ are not equal on certain direct sums of boxes}

We now introduce definition of {\it direct sum} of hypergraphs and boxes.

{\definition For two hypergraphs $G_1 = (V_{G_1},E_{G_1})$ and  $G_2 = (V_{G_2},E_{G_2})$, a direct sum of $G_1$ and $G_2$ is
$G_1\oplus G_2:= (V_{G_1\oplus G_2},E_{G_1\oplus G_2})$ with $V_{G_1\oplus G_2} = V_{G_1} \cup V_{G_2}$ and $E_{G_1\oplus G_2} = E_{G_1} \cup E_{G_2}$.
For any two boxes $B_1 = \{g(\lambda_c)\}_{c \in E_{G_1}}$ and $B_2 = \{g(\lambda_{c'})\}_{c' \in E_{G_2}}$ compatible with hypergraphs $G_1$ and $G_2$ respectively, their direct sum is a box $B_1\oplus B_2 := \{g(\lambda_c)\}_{c \in E_{G_1\oplus G_2}}$.
}

In the next part of this section, we use the following notation. By $p(\lambda)[V]$ we mean any joint probability distribution of the outputs of observables from set $V$, and by $p(\lambda)|_{V}$ the marginal probability distribution of $p(\lambda)$, defined on the outputs of observables of set $V$.
Moreover, by $D(g(\lambda_c)||p(\lambda_c))\left|_{p(\lambda)}\right.$
we mean the relative entropy distance between distribution of the output of variables from context $c$ of box $\{g(\lambda_c)\}$ and that from context $c$ of  non-contextual box defined by distribution $p(\lambda)$.

We will now need a lemma, which simplifies computation of $X_u$ and $X_{max}$ of direct sum of boxes, as it states, that it is enough to take minimization in both quantities only over product distributions.

{\lemma For any two hypergraphs $G_1=(V_{G_1},E_{G_1})$ and $G_2=(V_{G_2},E_{G_2})$ and boxes $B_1 \in C^{(n_1)}_{G_1}$ and $B_2 \in C^{(n_2)}_{G_2}$ there holds:
\beq
X_u(B_1\oplus B_2) 
 = \min_{p(\lambda)[V_{G_1}]p(\lambda)[V_{G_2}]} \sum_{c \in E_{G_1\oplus G_2}} {1\over n_1 + n_2} D(g(\lambda_c)||p(\lambda_c)),
\label{eq:xu-prod}
\eeq
and
\beq
X_{max}(B_1\oplus B_2) 
= \sup_{\{p(c)\}}\min_{p(\lambda)[V_{G_1}]p(\lambda)[V_{G_2}]} \sum_{c \in E_{G_1\oplus G_2}} p(c) D(g(\lambda_c)||p(\lambda_c)).
\label{eq:xmax-prod}
\eeq
\label{lem:xu-xmax-opt-prod}
}

{\it Proof.}

To see both equalities we observe that for any distribution $\{p(c)\}$ and any distribution $p(\lambda)[V_{G_1\oplus G_2}]$, there holds
\begin{multline}
\sum_{c \in E_{G_1} \cup E_{G_2}} p(c) D(g(\lambda_c)||p(\lambda_c))\left|_{p(\lambda)[V_{G_1\oplus G_2}]}\right. \\
= \sum_{c \in E_{G_1}} p(c) D(g(\lambda_c)||p(\lambda_c))\left|_{p(\lambda)[V_{G_1\oplus G_2}]|_{V_{G_1}}}\right. 
+\sum_{c \in E_{G_2}} p(c) D(g(\lambda_c)||p(\lambda_c))\left|_{p(\lambda)[V_{G_1\oplus G_2}]|_{V_{G_2}}}\right.  \\
= \sum_{c \in E_{G_1} \cup E_{G_2}} p(c) 
D(g(\lambda_c)||p(\lambda_c))|_{p(\lambda)[V_{G_1\oplus G_2}]|_{V_{G_1}}p(\lambda)[V_{G_1\oplus G_2}]|_{V_{G_2}}}
\end{multline}
This is because by definition of
$B_1\oplus B_2$ contexts from $E_{G_1}$ depend only on variables from $V_{G_1}$, similarly as contexts from $E_{G_2}$ depend only on $V_{G_2}$.
Hence $X_{\{p(c)\}}(B_1\oplus B_2) = \min_{p(\lambda)[V_{G_1}]p(\lambda)[V_{G_2}]} \sum_{c \in E_{G_1} \cup E_{G_2}}  D(g(\lambda_c)||p(\lambda_c))$, which
for $p(c) = {1\over n_1 + n_2}$ for all $c$ implies (\ref{eq:xu-prod}). Taking supremum over $\{p(c)\}$, we obtain (\ref{eq:xmax-prod}).\blacksquare

We are ready to show our main tool, interesting on its own, which is the following theorem that expresses $X_u$ and $X_{max}$ of a direct sum of two boxes in terms of these functions of these boxes.

{\theorem For any two hypergraphs $G_1$ and $G_2$ and boxes $B_1 \in C^{(n_1)}_{G_1}$ and $B_2 \in C^{(n_2)}_{G_2}$ there holds:
\be
X_u(B_1\oplus B_2) = {n_1 \over {n_1 + n_2}} X_u(B_1) + {n_2  \over {n_1 + n_2}} X_u(B_2),
\label{eq:xu-oplus-sum}
\ee
and
\be
X_{max}(B_1\oplus B_2) = max \{ X_{max}(B_1), X_{max}(B_2) \}.
\label{eq:xmax-oplus-max}
\ee
\label{thm:direct-sum}
}

{\it Proof.}

To see both the above statements, we observe that for a distribution $\{p(c)\}$ such that $ w = \sum_{c \in E_{G_1}} p(c) \neq 0$ and $w \neq 1$, and for any two distributions $p(\lambda)[V_{G_1}]$ and
$p(\lambda)[V_{G_2}]$, we have
\begin{multline}
\sum_{c \in E_{G_1} \cup E_{G_2}} p(c) D(g(\lambda_c)||p(\lambda_c))\left|_{p(\lambda)[V_{G_1}]p(\lambda)[V_{G_2}]}\right.  \\
= w \sum_{c \in E_{G_1}} {p(c)\over w} D(g(\lambda_c)||p(\lambda_c))\left|_{p(\lambda)[V_{G_1}]}\right.
+(1-w)\sum_{c \in E_{G_2}} {p(c)\over {1- w}} D(g(\lambda_c)||p(\lambda_c))\left|_{p(\lambda)[V_{G_2}]}\right.
\end{multline}
This immediately gives:
\begin{multline}
\min_{p(\lambda)[V_{G_1}]p(\lambda)[V_{G_2}]} \sum_{c \in E_{G_1} \cup E_{G_2}} p(c) D(g(\lambda_c)||p(\lambda_c))\left|_{p(\lambda)[V_{G_1}]p(\lambda)[V_{G_2}]}\right.  =\\
=w \min_{p(\lambda)[V_{G_1}]}\sum_{c \in E_{G_1}} {p(c)\over w} D(g(\lambda_c)||p(\lambda_c))
+(1-w)\min_{p(\lambda)[V_{G_2}]} \sum_{c \in E_{G_2}} {p(c)\over {1- w}} D(g(\lambda_c)||p(\lambda_c)).
\label{eq:w-minimas}
\end{multline}
Substituting $p(c) = {1\over n_1 + n_2}$ in the above equation, we obtain by lemma \ref{lem:xu-xmax-opt-prod} that
\begin{multline}
X_u(B_1\oplus B_2)
={n_1\over {n_1 + n_2}}[\min_{p(\lambda)[V_{G_1}]} {1\over n_1 } \sum_{c \in E_{G_1}} D(g(\lambda_c)||p(\lambda_c))] +
{n_2\over {n_1 + n_2}}[\min_{p(\lambda)[V_{G_2}]} {1\over n_2} \sum_{c \in E_{G_2}} D(g(\lambda_c)||p(\lambda_c))] \\
= {n_1\over {n_1 + n_2}} X_{u}(B_1) + {n_2\over {n_1 + n_2}}X_u(B_2),
\end{multline}
which proves the statement (\ref{eq:xu-oplus-sum}).

We now pass to prove the statement (\ref{eq:xmax-oplus-max}). We can assume w.l.g. that $X_{max}(B_1) \geq X_{max}(B_2)$.
By definition of $X_{max}$, for any $\delta_k > 0$, there exists $\{p_k(c)\}$ such that $X_{max}(B_1\oplus B_2) \leq X_{\{p_k(c)\}}(B_1\oplus B_2) + \delta_k$.
We will argue now, that for any $k$ and decreasing $\delta_k$, there holds $X_{max}(B_1\oplus B_2) \leq \max\{X_{max}(B_1),X_{max}(B_2)\} + \delta_k$, which will prove desired upper bound in limit $k \rightarrow \infty$. The proof then follows from the fact, that this upper bound can be attained by taking $w_k = \sum_{c \in E_{G_1}} p_k(c) = 1$ and such $\{p_k(c)\}$ that attain supremum in $X_{max}(B_1)$ in limit of large $k$.

We need to consider only two cases: $\{p_k(c)\}$ is such that $w_k = 1$ (case 1) or $0< w_k <1$ (case 2). In the first case we have $X_{max}(B_1\oplus B_2) \leq X_{\{p_k(c)\}}(B_1) +\delta_k \leq X_{max}(B_1) +\delta_k \leq \max \{X_{max}(B_1),X_{max}(B_2)\} + \delta_k$ which we aimed to prove. Thus, it is enough to show that in the second case $X_{max}(B_1\oplus B_2)$ is upper bounded by $X_{max}(B_1) +\delta_k$, since then the first case yields optimal value of $X_{max}(B_1\oplus B_2)$. Suppose then, that $\{p_k(c)\}$ satisfies $0 < w_k = \sum_{c \in E_{G_1}} p_k(c) < 1$.
This implies that
$\{{p_k(c)\over w_k}\}_{c\in E_{G_1}}$ and  $\{{p_k(c)\over 1- w_k}\}_{c\in E_{G_2}}$ are valid distributions, hence from (\ref{eq:w-minimas}), by lemma \ref{lem:xu-xmax-opt-prod}, we have
\beq
X_{max}(B_1\oplus B_2) \leq
w_k X_{\{{p_k(c)\over w_k}\}_{c\in E_{G_1}}}(B_1) + (1-w_k) X_{\{{p_k(c)\over 1- w_k}\}_{c\in E_{G_2}}}(B_2) + \delta_k.
\eeq
By definition of $X_{max}$ we have $X_{\{{p_k(c)\over w_k}\}_{c\in E_{G_1}}}(B_1)\leq X_{max}(B_1)$ and $X_{\{{p_k(c)\over 1- w_k}\}_{c\in E_{G_2}}}(B_2) \leq X_{max}(B_2)$ which gives from the above equality
\begin{multline}
X_{max}(B_1\oplus B_2)
\leq w_k X_{max}(B_1) + (1-w_k) X_{max}(B_2) +\delta_k \\
\leq X_{max}(B_1)+\delta_k= \max \{X_{max}(B_1),X_{max}(B_2)\} + \delta_k,
\end{multline}
hence, as we explained, the assertion follows.\blacksquare

The above theorem can be easily generalized to any finite direct sum of boxes, giving that $X_u$ is the average value of the $X_u$ of particular boxes from the direct sum (with weights according to cardinality of their number of contexts), and $X_{max}$ is the maximal value of $X_{max}$ on particular boxes.
We can state now the main application of this theorem.

{\corollary For any two hypergraphs $G_1$ and $G_2$, and a boxes $B_1 \in C^{(n_1)}_{G_1}$ and $B_2 \in C^{(n_2)}_{G_2}$ with $n_1, n_2\geq 1$, such that
$X_u(B_1) \neq X_u(B_2)$,  there holds
\be
X_{u}(B_1\oplus B_2)  < X_{max} (B_1\oplus B_2).
\ee
\label{cor:xu-diff}
}

{\it Proof}.

Since $X_u(B_1) \neq X_u(B_2)$, we can w.l.g. assume $X_u(B_1) > X_u(B_2)$. This implies, by theorem \ref{thm:direct-sum},
\begin{multline}
X_u(B_1\oplus B_2) = {n_1\over n_1 + n_2}X_u(B_1) + {n_2\over n_1 + n_2}X_u(B_2) < \\
X_u(B_1) \leq X_{max}(B_1) \leq \max\{X_{max}(B_1),X_{max}(B_2)\}
= X_{max}(B_1\oplus B_2),
\end{multline}
which proves the corollary.\blacksquare

From the above corollary we obtain immediately another one:

{\corollary For any two hypergraphs $G_1$ and $G_2$, a contextual box $B \in C^{(n_1)}_{G_1}$ and a non-contextual box $B_{nc} \in C^{(n_2)}_{G_2}$ with $n_1,n_2 \geq 1$, there holds
\beq
X_{u}(B\oplus B_{nc}) = {n_1 \over {n_1 + n_2}} X_{u}(B) <
X_{max} (B\oplus B_{nc}) = X_{max}(B).
\eeq
\label{cor:xu-c-nc-differ}
}
{\it Proof}.

It is enough to observe, that $X_u$ is faithful, hence $X_u(B) > X_u(B_{nc})$ and the corollary \ref{cor:xu-diff} applies.\blacksquare

The above corollary states that $X_{u}$ and $X_{max}$ differ on certain direct sums of boxes. Exemplary can be $PR\oplus PR_{1\over 2}$, since $PR_{1\over 2}$ is maximally mixed box, which is clearly non-contextual and $PR$ is contextual. There are also quantum boxes, i.e. that originate from performing certain measurements on a quantum state. Exemplary is defined as follows:

Consider the maximally entangled state $|\Psi^-\rangle=\frac{1}{\sqrt{2}}(|0\rangle_1|1\rangle_2-|1\rangle_1|0\rangle_2)$ and consider the hypergraph with $G:=(V_G, E_G)$ where $V_G=\{X_1, Z_1,$ ${-Z_2-X_2\over \sqrt2},$ ${Z_2-X_2\over \sqrt2}$, $R(X_1) \ot R(X_2)$, $R(Z_1) \ot R(Z_2)$ \} and $E_G:=\{ \{X_1, {-Z_2-X_2\over \sqrt2}\}, \{X_1, {Z_2-X_2\over \sqrt2}\}, \{Z_1, {-Z_2-X_2\over \sqrt2}\}, \{Z_1, {Z_2-X_2\over \sqrt2}\},$ $\{ R(X_1) \ot R(X_2),$ $R(Z_1) \ot R(Z_2)\} \}$. $X$ and $Z$ are Pauli matrices and $R(.)$ represents the rotation of Pauli matrix around $y$ axis by angle $\pi / 8$. We then consider a box $B$ obtained via measuring observables from $V_G$ on the state $|\Psi^-\>$ in groups defined by contexts. It is easy to see, that this box is a direct sum of the most non-local quantum box with two binary inputs and two binary outputs defined by first 4 observables and first 4 contexts, and a box with a single context, which is by definition non-contextual.

\subsection{KCBS box}
\label{app:computing:K}

A given hypergraph $G$ yields a definition of a large family of consistent boxes $C^{(n)}_G$. E.g. one can consider only those boxes which emerges as measurements of some observables on quantum state. One can narrow the latter family even more, by considering
in $V_G$ only observables that are rank-1 projectors (see e.g. \cite{YuOh,Cabello-13min,Megill-generation} and references therein). Then commensurability of 2 projectors turns to
be mutual orthogonality, and the hypergraph can be interpreted as {\it orthogonality graph}. This is the case for the Klyachko et al. result \cite{Klyachko-5}, where  a quantum state of qutrit is found, and measurements which give rise to maximal violation of the so called pentagon inequality. Such a pair: a quantum state and the set of measurements defines naturally a box called further KCBS box, denoted as K. Using similar techniques to that for xor-boxes, we find now the value of $X_u$ and $X_{max}$ for this box.

To be more precise the measurements given in \cite{Klyachko-5} form the set of 5 projectors $\{P_1,...,P_5\}$, designed in a such way, that they violate the so called pentagram inequality. This setup corresponds to a hypergraph that is pentagon, so that projectors commute in pairs: $P_1$ with $P_2$, $P_2$ with $P_3$ in a circle so that the last commutation is $P_5$ with $P_1$, so the graph is $G_{CH}^{(5)}$, but there are additional restrictions, since observables are one-dimensional projectors, so that their commutation implies their orthogonality. This implies that, e.g. the probability of the result $11$ which corresponds to obtaining an outcome 1 for both projectors as a result of measurement, is zero. More specifically, the Klyachko box is given by the same 5 distributions of the form $g(00)=1-\frac{2}{\sqrt{5}}$, $g(01)=g(10)=\frac{1}{\sqrt{5}}$ and $g(11)=0$. This implies that there are less symmetries than in the xor-boxes. After corresponding twirling, which we show below, there are 2 parameters left. Fortunately, we can make use of the fact that $g(11)=0$ in different way: it means that
the corresponding classical probability distribution should have $p(00)=0$ for all contexts, since the formula minimizes over the classical probabilities.

\begin{figure}
  \includegraphics[width=0.4\textwidth]{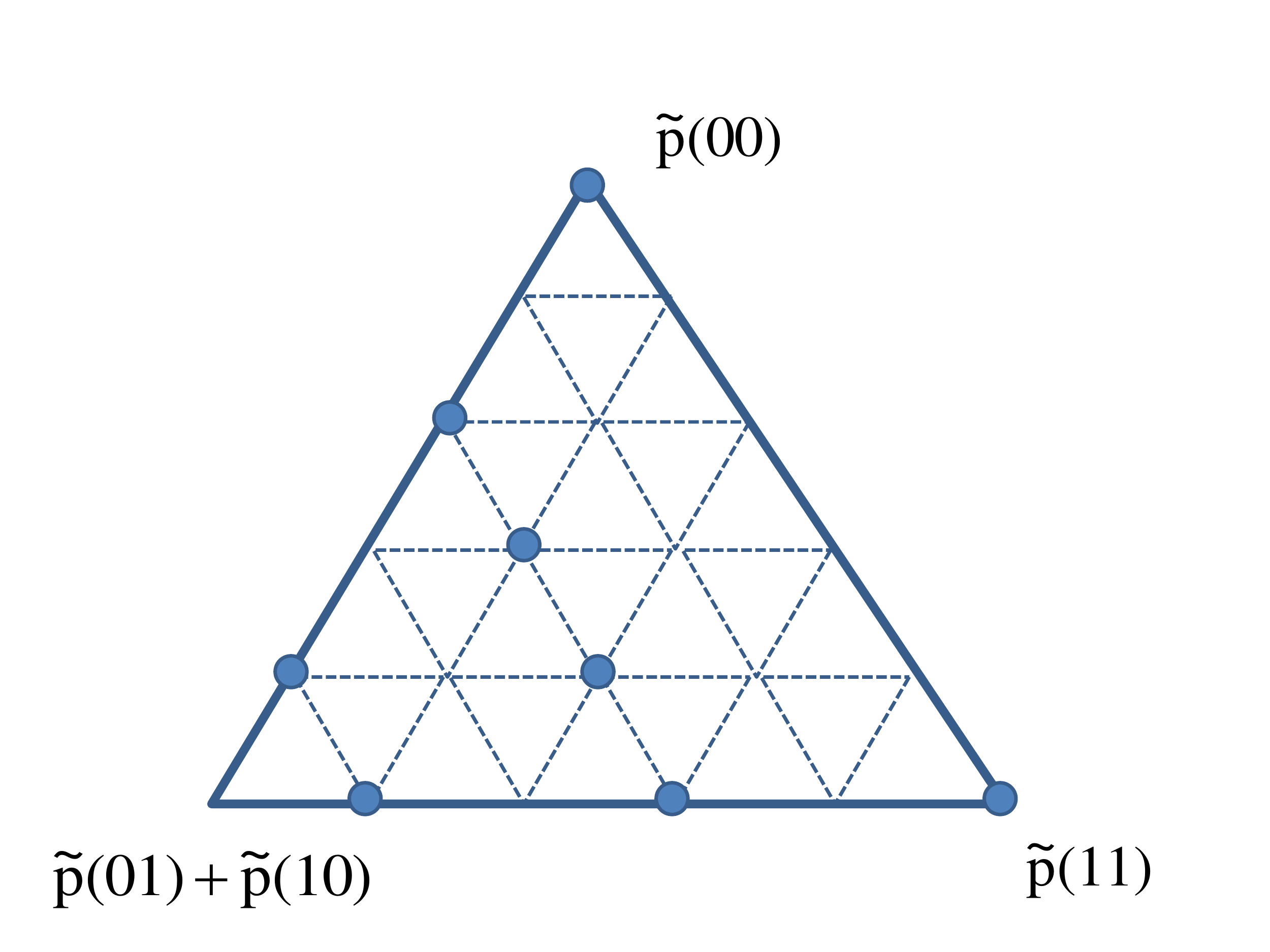}
  \caption{{\small The set of non-contextual distributions compatible with the Klyachko box, after twirling is situated within the trapezoid formed by the triangle without the vertex $\tilde{p}(01) + \tilde{p}(10)$. The bold points denote orbits, i.e. subsets of 32 extremal points invariant under $D_5$.}}\label{triangle}
\end{figure}

To describe the twirling consider the set of joined probability distributions $p(\lambda)$ with its $2^5=32$ extremal points. As in case of xor-boxes, we apply twirling determined by group $\mathcal{F}_{\mathcal{H}}$ which turns to be dihedral group $D_5$ consisting of  $d=10$ permutations. Due to symmetrization the marginal probability distributions calculated for the extremal points turns out to be context independent $p(\lambda_c)=\tilde{p}$ and to posses additional symmetry $\tilde{p}(01)=\tilde{p}(10)$.   As a consequence $32$ extremal points, which under the action of group $D_5$ form $8$ orbits (subsets of 32 extremal points invariant under $D_5$) the box can be characterized by only two parameters, e.g. $\tilde{p}(00)$, and $\tilde{p}(11)$ and conveniently visualized  by a $8$ points on a triangle plot as it is shown in Fig.\ref{triangle}. Thus a set of all symmetrized non-contextual distributions is a convex combination of $4$ distributions, namely
$\tilde{p}(00)=1$,
$\tilde{p}(11)=1$,
$\tilde{p}(00)=\frac{1}{5}$ and $\tilde{p}(10)=\tilde{p}(01)=\frac25$,
and finally
$\tilde{p}(11)=\frac{1}{5}$ and $\tilde{p}(10)=\tilde{p}(01)=\frac25$.
Now our goal is to find minimum in $X_u(K)$ which, due to all mentioned symmetries, is given simply by
\be
X_{u}(K)=\min_{\tilde{p}} D(g(\lambda_{c_0})||\tilde{p}).
\ee
Note that $g(11)=0$, so that looking for a minimum we can restrict to the case of $\tilde{p}(11)=0$. In this way the problem of calculating $X_u$ has been reduced to finding minimum over a single parameter $\tilde{p}(00)$ in the range between $0.2$ and $1$
\be
X_{u}(K)=\min_{0.2\leq\tilde{p}(00)\leq 1}  \chi(g(00),\tilde{p}(00)),
\ee
where
\be
\chi(x,y)=x \log\Big(\frac{x}{y}\Big)+(1-x) \log\Big(\frac{1-x}{1-y}\Big).
\ee
$\chi(x,y)$ is strictly increasing function of argument $y$ provided that
$0\leq x \leq y \leq 1$ which is seen from equation
\be
\frac{\partial \chi(x,y)}{\partial y}=\frac{y-x}{y(1-y)}.
\ee
So finally we get $X_u(K)=\chi(g(00),0.2)\approx0.0466576$. For reasons similar to that given for xor-boxes, we have in this case $X_{max}=X_u$.

\section{Additivity results}
\label{sec:app:add}

In this section we prove that for exemplary xor-boxes $X_u$ is additive, and that it is $2$-copy additive for isotropic xor-boxes considered in this paper.
We begin with Definition and necessary lemmas. The main results are theorems \ref{thm:2add} and \ref{thm:ex-add}, and their main application is stated in Corollary \ref{cor:iso-2add}.

{\definition For any two hypergraphs $G_1=(V_{G_1},E_{G_1})$ and $G_2=(V_{G_2},E_{G_2})$, we define their tensor product to be the hypergraph
\be
G_1 \otimes G_2:=(V_{G_1 \otimes G_2}, E_{G_1 \otimes G_2})
\ee
where $V_{G_1 \ot G_2}:=(V_{G_1}\cup V_{G_2})$ and $E_{G_1 \ot G_2}:=\{c \cup c'|c \in E_{G_1}$ and $c' \in E_{G_2} \}$. For two boxes $B_1=\{g_1(\lambda_c)\}$ and $B_2=\{g_2(\lambda_{c'})\}$ compatible with hypergrahps $G_1$ and $G_2$ respectively, their tensor product is a box compatible with $G_1\ot G_2$ given by $B_1\ot B_2 := \{g_1(\lambda_c)g_2(\lambda_{c'})\}$ i.e. such that the distribution of its context $c \cup c'$ is a product of distributions $g_1(\lambda_c)$ and $g_2(\lambda_{c'})$.
\label{def:tensor}
}

We make now an observation, which characterizes the set of noncontextual boxes of the tensor product of two the same hypergraphs.

{\observation The set of noncontextual boxes $NC_{G^{\otimes 2}}$ belonging to $C^{(n)}_G\ot C^{(n)}_G$ is spanned by tensor products of extremal points of the set $NC_G$.
\label{obs:nonc}}

{\it Proof.}

To see this, consider an extremal point of $NC_{G^{\otimes 2}}$. It is equal to a box with joint distribution over $2n$ observables $\delta_{{\bf a, a_0}}$ for some ${\bf a_0}$. Such a distribution is a product of distributions $\delta_{{\bf a_1, a_{01}}}$ and $\delta_{{\bf a_2, a_{02}}}$ where ${\bf a_1}$ and ${\bf a_2}$ are output strings of outputs $a_i$ and each $a_i \in \{0,1,....,d_{A_i}\}$. ${\bf a_{01}}$ and ${\bf a_{02}}$ are some fixed output strings. ${\bf a_1}$ and ${\bf a_2}$ when written in a system with basis $d$ (assuming that all of them are equal, otherwise one has to consider a multibase system) and concatenating yields ${\bf a}$. Hence any extremal point of $NC_{G^{\otimes 2}}$ is a product of extremal points of the set $NC_G$.\blacksquare

We will need also a lemma stated in general for linear operations, which will be used for twirling operation:

{\lemma After any linear operation $\tau$ on any convex set $Y$, the set of extremal points of the image set $\tau(Y)$ is the subset of the set of images of extremal points of $Y$ through $\tau$.
\label{lem:tau}
}

{\it Proof.}

Consider any point $\tau(B)$ which is an image of non-extremal point $B$ in $Y$. Since $\tau$ is linear, we have $\tau(B)=\tau(p_1 B_1 + (1-p_1) B_2) = p_1\tau(B_1)+ (1-p_1)\tau(B_2)$
hence it cannot be extremal in $\tau(Y)$. Thus, any extremal point in $\tau(Y)$ must be an image of extremal point in $Y$. \blacksquare

This enables us to state the following observation:

{\observation For any linear map $\tau : C^{(n)}_G \rightarrow C^{(n)}_G$ the set $\tau \ot \tau (NC_{G^{\otimes 2}})$ is spanned by tensor products of extremal points of the set $\tau(NC_G)$
\label{lem:span}}

{\it Proof.}

By lemma \ref{lem:tau} the only extremal points in $\tau\ot\tau(NC_{G^{\ot 2}})$ are within the set of images of extremal points through $\tau$. We know from observation \ref{obs:nonc} that extremal points of $\tau\ot \tau(NG_{G^{\otimes 2 }})$ are of the form $\tau(E_i)\ot \tau(E_j)$ where $E_k$ are extremal points of $NC_G$. Now if $\tau(E_i)$ is not extremal in $\tau(NC_G)$ i.e. can be decomposed into $\sum_i p_i \tau(E_{k_i})$ then clearly the image $\tau\ot\tau(E_i\ot E_j)$ for any $j$ is not an extremal in $\tau\ot\tau(NC_{G^{\ot 2}})$, as it can be decomposed into nontrivial mixture $\sum_i p_i \tau(E_{k_i}) \ot \tau(E_j)$. The same argument holds for $\tau(E_j)$ : it cannot be non-extremal in $\tau(NC_G)$ if the pair $\tau(E_i)\ot \tau(E_j)$ is extremal in
$\tau\ot\tau(NC_{G^{\ot 2}})$. Hence, the only extremal points in $\tau\ot\tau(NC_{G^{\ot 2}})$ are the tensor products of extremal points in $\tau(NC_G)$.\blacksquare

In what follows, for two arbitrary boxes $B_1$ and $B_2$ by interval $[B_1,B_2]$ we mean the set $\{ pB_1 +(1-p)B_2| p\in[0,1]\}$.

{\lemma {\label {obs:single_term2}} Let box $B=\{g(\lambda_c)\} \in C^{(n)}_G$ be invariant under some linear operation $\tau$, which maps all boxes on $C^{(n)}_G$ into interval $[B_e,B_e']$ and maps $NC_G$ into interval $[L,L'] \subset [B_e,B_e']$. Let also some of $g(\lambda_c)$ be equal to $g(\lambda_{c_0})$ and the rest of the $g(\lambda_c)$ be equal to $\Pi(g(\lambda_{c_0}))$ for some reversible operation $\Pi$. Then, there holds:
\be
X_u(B^{\ot 2})=\inf_{P_{nc}\in\tau\ot\tau(NC_{G^{\ot 2}})} D(g(\lambda_{c_0})g(\lambda_{c_0})||p_{\bar 1})
\ee
where $g(\lambda_{c_0})g(\lambda_{c_0})$ is a product of distributions $g(\lambda_{c_0})$ and $p_{\bar 1}$ is the distribution of some fixed context number $\bar 1$ of $P_{nc}$.}

{\it Proof:}
Let $n_1$ be the number of the contexts of $B$ with the same distribution $g(\lambda_{c_0})$ and $n_2$ the number of the remaining contexts with distribution $\Pi(g(\lambda_{c_0}))$.
In what follows, we identify $g(\lambda_{c_0})$ with $q$ and $\Pi(g(\lambda_{c_0}))$ with $\bar{q}$ for short.
We know that
\be
X_u(B^{\ot 2}) = \inf_{P_{nc} \in \tau\ot\tau(NC_{G^{\ot 2}})}{1\over n^2}\sum_{c,c'} D(g(\lambda_c)g(\lambda_{c'}) || p_{cc'})
\ee
where $n^2$ is total no. of contexts and $P_{nc}=\{p_{cc'}\}$. From the Observations {\ref{obs:nonc}} and {\ref{lem:span}}, the box $P_{nc}$ can be written as
\be
P_{nc}= p_1 L{\tilde L} + p_2 L{\tilde B_e'} + p_3 B_e'{\tilde L} +p_4 B_e'{\tilde B_e'}.
\ee
We switch now from equality for boxes to equality for contexts, using for short the notation
$B_e=\{e_c\}$ meaning that $e_c$ is the context number $c$ of a box $B$ and  similarly $B_e'=\{e_c'\}$ and $L=\{l_c\}$, $L'=\{e_c'\}$. The above equality gives for each $c$ and $c'$:
\be
p_{cc'} = p_1 l_c l_{c'} + p_2 l_c e_{c'}'+ p_3 e_c' l_{c'} + p_4 e_c' e_{c'}'
\ee
Now, consider the following 4 cases, where due to $[L,L']\subset [B_e,B_{e'}]$ we can set $L = s B_e + (1-s) B_e'$ for some $s\in[0,1]$.

{\bf Case 1.} $\forall_{ c\in\{n_1\}, c'\in\{n_1\}}$
\begin{multline}
D(g(\lambda_c)g(\lambda_{c'})||p_{cc'}) =
D(qq||p_1 (sq+(1-s){\bar q})(sq+(1-s){\bar q})
+ p_2 (sq+(1-s){\bar q}) {\bar q} \\\newline
+ p_3 {\bar q}(sq+(1-s){\bar q}) + p_4 {\bar q}{\bar q})
\label{eq:re-first}
\end{multline}

{\bf Case 2.} $\forall_{ c\in\{n_2\}, c'\in\{n_1\}}$
\begin{multline}
D(g(\lambda_c)g(\lambda_{c'})||p_{cc'}) =
D({\bar q}q||p_1 (s{\bar q}+(1-s)q)(sq+(1-s){\bar q})
+ p_2 (s{\bar q}+(1-s)q) {\bar q} \\\newline
+ p_3 q(sq+(1-s){\bar q}) + p_4 q{\bar q})
\end{multline}
Applying reversible operations $\Pi_{cc'}= \Pi \ot I$ to both distributions in the above relative entropy term, we get
\beq
D(qq||p_1 (sq+(1-s){\bar q})(sq+(1-s){\bar q})
+ p_2 (sq+(1-s){\bar q}) {\bar q} 
+ p_3 {\bar q}(sq+(1-s){\bar q}) + p_4 {\bar q}{\bar q})
\eeq
which is exactly relative entropy term in (\ref{eq:re-first}).
Similarly, by considering other two cases where $c\in \{n_1\}\, \&\, c'\in \{n_2\}$ and $c\in \{n_2\}\, \&\, c'\in \{n_2\}$ we get the same equality after applying reversible operations $\Pi_{cc'}=I\ot \Pi$ and $\Pi_{cc'}=\Pi\ot \Pi$ respectively, and the assertion follows $\blacksquare$

{\observation In general lemma \ref{obs:single_term2} holds for n-copy, i.e.
\be
X_u(B^{\ot n})=\inf_{P_{nc}\in\tau\ot...\ot\tau(NC_{G^{\ot n}})} D(g(\lambda_{c_0})g(\lambda_{c_0})....||p_{\bar 1})
\ee
\label{obs:n-one-term}
}

{\it Proof.}

The proof goes in full analogy to that of lemma \ref{obs:single_term2}.\blacksquare

We can state now one of the main theorems of this section.

{\theorem Let box $B =\{g(\lambda_c)\} \in C^{(n)}_G$ and let the image of $C^{(n)}_G$ through $\tau^{{\mathcal L}_0}_B$ be the interval $[B_e,B_e']$ and the image of set $NC_G$ be $[L,L'] \subset [B_e,B_e']$ such that $L = s B_e + (1-s) B_e'$ with $s > \frac{1}{2}$ and $B = rB_e +(1-r)B_{e'}$ with $r>s$. Let also $B_e=\{e_c\}$ and $B_e'=\{e_c'\}$, such that $e_c$ has disjoint support from $e_c'$, then there holds
\be
X_u(B\otimes B) = 2 X_u(B).
\ee
\label{thm:2add}
}

{\it Proof}.

We first note that by theorem \ref{thm:iso-optimal}, with the set of automorphisms $\mathcal{L}'_0$ being the set of all tensor products of automorphisms from $\mathcal{L}_0$ with themselves, we have:
\be
X_u(B^{\ot 2}) = \inf_{P_{nc}\in \tau\ot\tau(NC_{G^{\ot 2}})} {1\over n^2}\sum_{c,c'} D(g(\lambda_c)g(\lambda_{c'}) || p_{cc'})
\ee
Now, by lemma \ref{obs:single_term2}, we have
\be
X_u(B^{\ot 2})=\inf_{P_{nc}\in \tau\ot\tau(NC_{G^{\ot 2}})} D(q q||p_{ij})
\ee
where $q= r e_i + (1-r) e_i'$ (also $q= r e_j + (1-r) e_j'$ ) ($r>s$ by assumption) and indices $i, j$ represent a fixed context of $P_{nc}$ such that all distributions of $P_{nc}$ are transformable into it, by operations which at the same time transform all distributions of $B^{\ot 2}$ into $qq$.
By theorem \ref{thm:iso-optimal} and using the fact that $qq$ is invariant under swap (it can be achieved by local or global swap operations depending on the hypergraph under consideration), it is equivalent to the quantity:
\beq
X_u(B^{\ot 2})=\inf_{p_1+p_2+p_3=1} D(q q|| p_1 l_i l_j 
+ {p_2\over 2}(l_i l_j' + l_i' l_j) + p_3 l_i' l_j')
\eeq
Note, that we can relax the minimization and hence we have the following lower bound:
\beq \label {eq:2copy}
X_u(B^{\ot 2})\geq \inf_{p_1+p_2+p_3=1} D(q q|| p_1 l_i l_j 
+ {p_2\over 2}(l_i e_j' + e_i' l_j) + p_3 e_i' e_j')
\eeq
if we are lucky to find the solution that is non-contextual, then we will find solution to our initial minimization problem. As we will see,
this will be the case.

Using the fact that $e_i (e_j)$ and $e_i' (e_j')$ have disjoint supports,  decomposing $l_i = s e_i + (1-s) e_i'$ and $l_j = s e_j + (1-s) e_j'$ we get that:
\begin{multline}
X_u(B^{\ot 2}) \geq \inf_{p_1+p_2+p_3=1} \big[ r^2\sum_{\bf a} p^{({\bf a})}_{e_i e_j} \log ( {r^2 p^{({\bf a})}_{e_i e_j}\over (p_1 s^2) p^{({\bf a})}_{e_i e_j} }) 
+ 2r(1-r)\sum_{\bf a} p^{({\bf a})}_{e_i e_j'} \log ( {r(1-r)p^{({\bf a})}_{e_i e_j'} \over (p_1 s(1-s) + \frac{s p_2}{2})p^{({\bf a})}_{e_i e_j'} }) \\\newline
+ (r-1)^2 \sum_{\bf a} p^{({\bf a})}_{e_i' e_j'} \log ( {(r-1)^2p^{({\bf a})}_{e_i' e_j'} \over (p_1(1-s)^2 + p_2(1-s)
+ p_3)p^{({\bf a})}_{e_i 'e_j'} }) \big ]
\end{multline}
where $\{p^{({\bf a})}_{e_i e_j}\}$ is the distribution of $e_i e_j$.
For $r=1$ i.e. when $q = e_i$ which is the case for PR-box, PM-box, Mermin's star and CH-box, we have that
\ben
X_u(B^{\ot 2}) \geq \inf_{p_1+p_2+p_3=1} \log {1 \over p_1 s^2}
\een
where LHS is clearly minimal for $p_1 = 1$, which means that the closest distribution in our set is non-contextual, equal to $l_i l_j$, hence
\be
X_u(B^{\ot 2}) = \log {1\over s^2}= 2 X_u(B)
\ee
Consider now $r< 1$. Here we are able to prove additivity for 2 copies, by using Lagrange multipliers approach. We need to find infimum of
\beq
X_u(B^{\ot 2}) \geq \inf_{p_1+p_2+p_3=1} \big [r^2\log ( {r^2\over p_1 s^2} ) 
+ 2r(1-r)\log ( {r(1-r) \over p_1 s(1-s) + \frac{s p_2}{2}) }
+ (r-1)^2 \log ( {(r-1)^2 \over p_1(1-s)^2 + p_2(1-s) + p_3 })]
\eeq

We first check if the infimum is attained in the interior of the simplex of the boundary conditions. Using Mathematica v07 we obtain, that there
is only 1 solution of the set of Lagrange equations:
\ben
p_1 = {r^2\over s^2}, \nonumber \\
p_2 = {2 r (-r + s)\over s^2}, \nonumber \\
p_3 = {(r - s)^2\over s^2}
\een
However, we have that $q$ is contextual, so $r > s$, which gives that $p_2$ of the above solution is negative. Hence the function does not have
infimum in the interior, in the considered region of parameters $r$ and $s$. It suffice to consider boundaries, i.e. cases $p_3 = 0$, $p_2 = 0$ and
$p_2 = p_3 = 0$ (other cases are excluded by the fact that $p_1 > 0$). Again, using Lagrange multipliers method, we solve the first two cases.
The first one has two solutions, which has $p_2 < 0$ if $s > {1\over 2}$. In case $p_3 = 0$ we observe that $p_1 > 1$, which finally proves that the
only solution that attributes to infimum is $p_1 =1$, which is non-contextual solution as in case of extremal q, that yields additivity i.e. $X_u(B^{\ot 2}) = 2 X_u(B)$.$\blacksquare$

{\theorem Under assumptions of theorem \ref{thm:2add}, $X_u$ is additive on $B_e$ and $B_e'$.
\label{thm:ex-add}
}

{\it Proof.}

In the calculation of relative entropy for this case, we will have similar terms as in eqn. (\ref{eq:2copy}) but with n copies. By observation \ref{obs:n-one-term}, we have:
\beq \label {eq:ncopy}
X_u(B^{\ot n})\geq \inf_{\sum p_i=1} D(qq...|| p_1 l_1 l_2...+T)
\eeq
where $T$ is all the other possible terms of $l_n$s \& $e_n$s with weights $p_n$. Note here that, $l_n$s \& $e_n$s are all some fixed context. Since, $l_n=s e_n+(1-s)e_n'$ for all $n$ we have,
\beq
X_u(B^{\ot n})\geq \inf_{\sum p_i=1} D( r^n e_1 e_2...+ T_1|| p_1 s^n e_1 e_2...+T_1')
\eeq
where $T_1$ and $T_1'$ are all the other possible terms of $e_n$s and $e_n'$s with $T_1$ having powers $(1-r)$ while $T_1'$ have different weights $p_i$.
\beq
X_u(B^{\ot n}) \geq \inf_{\sum p_{i=1}} {\sum_{\bf a}} r^n p_{e_1 e_2..}^{({\bf a})} \log ({r^n p^{({\bf a})}_{e_1 e_2..} \over p_1 s^n p^{({\bf a})}_{e_1 e_2..}}) +T_2
\eeq
where $T_2$ contains terms with powers of $(1-r)$. For extremal points $r=1$ therefore,
\be
X_u(B^{\ot n}) \geq \inf_{\sum p_{i=1}} {\sum_{\bf a}} p_{e_1 e_2..}^{({\bf a})} \log ({p^{({\bf a})}_{e_1 e_2..} \over p_1 s^n p^{({\bf a})}_{e_1 e_2..}})
\ee

\be
X_u(B^{\ot n})\geq \inf_{p_1} {\sum_{\bf a}} p_{e_1 e_2..}^{({\bf a})} \log ({1\over p_1 s^n})
\ee
Since ${\sum_{\bf a}} p_{e_1 e_2..}^{({\bf a})}=1$ and minimum is attained at $p_1=1$ which gives us desired proof
\be
X_u(B^{\ot n})\geq n\log\frac{1}{s}=nX_u(B)
\ee
Analogously we can show additivity of $X_u$ on $B_e'$ by exchanging $l_0$ to $l_0'$ and $e_0'$ to $e_0$.
$\blacksquare$
{\corollary {\label {cor:iso-2add}}For a box $B \in I^{\mathcal{L}_0}_{PM}\cup I^{\mathcal{L}_0}_M \cup I^{\mathcal{L}_0}_{CH_{(n)}}$ $X_u(B^{\ot 2})= 2 X_u(B)$. For $B \in PM \cup M \cup CH_{(n)}\cup PM' \cup M' \cup CH_{(n)}'$ $X_u$ is additive.
}

{\it Proof}.

To see the first statement, it suffices to check that the box $B \in I^{\mathcal{L}_0}_{PM}\cup I^{\mathcal{L}_0}_M \cup I^{\mathcal{L}_0}_{CH_{(n)}}$ satisfies assumptions of theorem \ref{thm:2add}. The second is direct result from theorem \ref{thm:ex-add}.$\blacksquare$

\end{appendix}

\end{document}